\documentclass[twocolumn,times]{aastex63}
\usepackage{amsmath}

\newcommand\degree{\hbox{$^\circ$}}

\newcommand\change[1]{#1}%\textcolor{red}{#1}}

\shorttitle{COMAP Early Science VI: COMAP GPS}
\shortauthors{Rennie et al.}
\graphicspath{{./}{}}
\NewPageAfterKeywords

\begin{document}

\title{COMAP Early Science: VI. A First Look at the COMAP Galactic Plane Survey}

\correspondingauthor{Thomas J. Rennie}
\email{thomas.rennie@manchester.ac.uk}

\author[0000-0002-1667-3897]{Thomas J. Rennie}
\affiliation{Jodrell Bank Centre for Astrophysics, Alan Turing Building, Department of Physics and Astronomy, School of Natural Sciences, The University of Manchester, Oxford Road, Manchester, M13 9PL, U.K.}

\author[0000-0001-7911-5553]{Stuart E.~Harper}
\affiliation{Jodrell Bank Centre for Astrophysics, Alan Turing Building, Department of Physics and Astronomy, School of Natural Sciences, The University of Manchester, Oxford Road, Manchester, M13 9PL, U.K.}

\author[0000-0002-0045-442X]{Clive Dickinson}
\affiliation{Jodrell Bank Centre for Astrophysics, Alan Turing Building, Department of Physics and Astronomy, School of Natural Sciences, The University of Manchester, Oxford Road, Manchester, M13 9PL, U.K.}

\author[0000-0001-7612-2379]{Liju Philip}
\affiliation{Jet Propulsion Laboratory, California Institute of Technology, 4800 Oak Grove Drive, Pasadena, CA 91109,  USA}

\author[0000-0002-8214-8265]{Kieran A. Cleary}
\affiliation{California Institute of Technology, 1200 E. California Blvd., Pasadena, CA 91125, USA}

% TIER II

\author[0000-0003-2358-9949]{Richard J. Bond}
\affiliation{Canadian Institute for Theoretical Astrophysics, University of Toronto, 60 St. George Street, Toronto, ON M5S 3H8, Canada}

\author{Jowita Borowska}
\affiliation{Institute of Theoretical Astrophysics, University of Oslo, P.O. Box 1029 Blindern, N-0315 Oslo, Norway}

\author[0000-0001-8382-5275]{Patrick C.~Breysse}
\affiliation{Center for Cosmology and Particle Physics, Department of Physics, New York University, 726 Broadway, New York, NY, 10003, USA}

\author{Morgan Catha}
\affil{Owens Valley Radio Observatory, California Institute of Technology, Big Pine, CA 93513, USA}

\author[0000-0002-9043-2645]{Roke Cepeda-Arroita}
\affiliation{Jodrell Bank Centre for Astrophysics, Alan Turing Building, Department of Physics and Astronomy, School of Natural Sciences, The University of Manchester, Oxford Road, Manchester, M13 9PL, U.K.}

\author[0000-0003-2618-6504]{Dongwoo T.~Chung}
\affiliation{Canadian Institute for Theoretical Astrophysics, University of Toronto, 60 St. George Street, Toronto, ON M5S 3H8, Canada}
\affiliation{Dunlap Institute for Astronomy and Astrophysics, University of Toronto, 50 St. George Street, Toronto, ON M5S 3H4, Canada}

\author[0000-0003-2358-9949]{Sarah E.~Church}
\affiliation{Kavli Institute for Particle Astrophysics and Cosmology \& Physics Department, Stanford University, Stanford, CA 94305, US}

\author[0000-0002-5223-8315]{Delaney A.~Dunne}
\affiliation{California Institute of Technology, 1200 E. California Blvd., Pasadena, CA 91125, USA}

\author[0000-0003-2332-5281]{Hans Kristian Eriksen}
\affiliation{Institute of Theoretical Astrophysics, University of Oslo, P.O. Box 1029 Blindern, N-0315 Oslo, Norway}

\author[0000-0001-8896-3159]{Marie Kristine Foss}
\affiliation{Institute of Theoretical Astrophysics, University of Oslo, P.O. Box 1029 Blindern, N-0315 Oslo, Norway}

\author{Todd Gaier}
\affiliation{Jet Propulsion Laboratory, California Institute of Technology, 4800 Oak Grove Drive, Pasadena, CA 91109, USA}

\author{Joshua Ott Gundersen}
\affiliation{Department of Physics, University of Miami, 1320 Campo Sano Avenue, Coral Gables, FL 33146, USA}

\author[0000-0001-6159-9174]{Andrew I.~Harris}
\affil{Department of Astronomy, University of Maryland, College Park, MD 20742}

\author[0000-0001-7449-4638]{Brandon Hensley}
\affiliation{Department of Astrophysical Sciences, Princeton University, Princeton, NJ 08544, USA}

\author{Richard Hobbs}
\affil{Owens Valley Radio Observatory, California Institute of Technology, Big Pine, CA 93513, USA}

\author[0000-0003-3420-7766]{H\aa vard T.~Ihle}
\affiliation{Institute of Theoretical Astrophysics, University of Oslo, P.O. Box 1029 Blindern, N-0315 Oslo, Norway}

\author[0000-0002-5959-1285]{James W.~Lamb}
\affil{Owens Valley Radio Observatory, California Institute of Technology, Big Pine, CA 93513, USA}

\author{Charles R.~Lawrence}
\affiliation{Jet Propulsion Laboratory, California Institute of Technology, 4800 Oak Grove Drive, Pasadena, CA 91109, USA}

\author{Jonas G.\ S.\ Lunde}
\affiliation{Institute of Theoretical Astrophysics, University of Oslo, P.O. Box 1029 Blindern, N-0315 Oslo, Norway}

\author[0000-0002-5158-243X]{Roberta Paladini}
\affiliation{Infrared Processing Analysis Center, California Institute of Technology, Pasadena, CA 91125, USA}

\author[0000-0001-5213-6231]{Timothy J.~Pearson}
\affiliation{California Institute of Technology, 1200 E. California Blvd., Pasadena, CA 91125, USA}

\author{Maren Rasmussen}
\affiliation{Institute of Theoretical Astrophysics, University of Oslo, P.O. Box 1029 Blindern, N-0315 Oslo, Norway}

\author[0000-0001-9152-961X]{Anthony C.S.~Readhead}
\affiliation{California Institute of Technology, 1200 E. California Blvd., Pasadena, CA 91125, USA}

\author[0000-0001-5301-1377]{Nils-Ole Stutzer}
\affiliation{Institute of Theoretical Astrophysics, University of Oslo, P.O. Box 1029 Blindern, N-0315 Oslo, Norway}

\author[0000-0002-5437-6121]{Duncan J.~Watts}
\affiliation{Institute of Theoretical Astrophysics, University of Oslo, P.O. Box 1029 Blindern, N-0315 Oslo, Norway}

\author[0000-0003-3821-7275]{Ingunn Kathrine Wehus}
\affiliation{Institute of Theoretical Astrophysics, University of Oslo, P.O. Box 1029 Blindern, N-0315 Oslo, Norway}

\author{David P.~Woody}
\affil{Owens Valley Radio Observatory, California Institute of Technology, Big Pine, CA 93513, USA}

\collaboration{42}{(COMAP Collaboration)}

%% Note that the \and command from previous versions of AASTeX is now
%% depreciated in this version as it is no longer necessary. AASTeX 
%% automatically takes care of all commas and "and"s between authors names.

%% AASTeX 6.31 has the new \collaboration and \nocollaboration commands to
%% provide the collaboration status of a group of authors. These commands 
%% can be used either before or after the list of corresponding authors. The
%% argument for \collaboration is the collaboration identifier. Authors are
%% encouraged to surround collaboration identifiers with ()s. The 
%% \nocollaboration command takes no argument and exists to indicate that
%% the nearby authors are not part of surrounding collaborations.

%% Mark off the abstract in the ``abstract'' environment. 
\begin{abstract}
We present early results from the COMAP Galactic Plane Survey conducted between June 2019 and April 2021, spanning $20^\circ<\ell<40^\circ$ in Galactic longitude and $|b|<1.\!\!^{\circ}5$ in Galactic latitude with an angular resolution of $4.5^{\prime}$. We present initial results from the first part of the survey, including diffuse emission and spectral energy distributions (SEDs) of H\textsc{ii} regions and supernova remnants. Using low and high frequency surveys to constrain free-free and thermal dust emission contributions, we find evidence of excess flux density at $30$\,GHz in six regions that we interpret as anomalous microwave emission. Furthermore we model UCH\textsc{ii} contributions using data from the $5$\,GHz CORNISH catalogue and reject this as the cause of the $30$\,GHz excess. Six known supernova remnants (SNR) are detected at $30$\,GHz, and we measure spectral indices consistent with the literature or show evidence of steepening. The flux density of the SNR W44 at $30$\,GHz is consistent with a power-law extrapolation from lower frequencies with no indication of spectral steepening in contrast with recent results from the Sardinia Radio Telescope. We also extract five hydrogen radio recombination lines (RRLs) to map the warm ionized gas, which can be used to estimate electron temperatures or to constrain continuum free-free emission. The full COMAP Galactic plane survey, to be released in 2023/2024, will \change{span $\ell \sim 20^{\circ}$--$220^{\circ}$ and will be the first large-scale radio continuum and RRL survey at $30$\,GHz with $4.5^{\prime}$ resolution.}

\end{abstract}

%% Keywords should appear after the \end{abstract} command. 
%% The AAS Journals now uses Unified Astronomy Thesaurus concepts:
%% https://astrothesaurus.org
%% You will be asked to selected these concepts during the submission process
%% but this old "keyword" functionality is maintained in case authors want
%% to include these concepts in their preprints.
%\keywords{surveys -- ISM: HII regions -- ISM: supernova remnants -- ISM: general: Galaxy: general -- radio continuum: ISM}

%% From the front matter, we move on to the body of the paper.
%% Sections are demarcated by \section and \subsection, respectively.
%% Observe the use of the LaTeX \label
%% command after the \subsection to give a symbolic KEY to the
%% subsection for cross-referencing in a \ref command.
%% You can use LaTeX's \ref and \label commands to keep track of
%% cross-references to sections, equations, tables, and figures.
%% That way, if you change the order of any elements, LaTeX will
%% automatically renumber them.
%%
%% We recommend that authors also use the natbib \citep
%% and \citet commands to identify citations.  The citations are
%% tied to the reference list via symbolic KEYs. The KEY corresponds
%% to the KEY in the \bibitem in the reference list below. 

% Introduction
\section{Introduction} \label{sec:intro}

Surveys of the Galaxy at radio frequencies ($\nu \lesssim 100$\,GHz) offer a largely unobstructed view of the interstellar medium. For example, observations of atomic cold gas via the 21\,cm (1.4\,GHz) H\textsc{i} line can map the gas throughout the entire Galactic disk. Furthermore, radio data provide unique information on the Galaxy not easily seen at \change{other} wavelengths such as the large radio loops \change{\citep{Dickinson2018,Planck2015XXV}}. Radio emission can be used to estimate star-formation rates as well as to study the diverse range of Galactic objects such as supernova remnants, H\textsc{ii} regions, planetary nebulae, and molecular \change{clouds} \citep[e.g.,][]{Brunthaler2021}.

At radio frequencies, continuum emission comes from three phases of the interstellar medium \citep{Draine2011}: i) synchrotron emission produced by relativistic cosmic rays (mostly electrons) accelerated by the Galactic magnetic field, ii) free-free emission from warm ($T_e \!\sim \!10^4$\,K) ionized gas, mostly around hot O/B stars, and iii) thermal vibrational (and spinning dust) emission \change{from cold} \change{($T_d\!\sim\!15$--20\,K)} dust grains. Observations of the radio continuum over a wide range of frequencies can be used to separate the various continuum emission mechanisms into their individual components---namely, synchrotron, free-free, spinning dust or anomalous microwave emission (AME), and thermal dust emission. This form of component separation has been an important aspect of cosmic microwave background (CMB) foreground removal, which typically uses the different frequency (or sometimes spatial) response of each component to allow them to be separated \citep{Leach2008,Dunkley2009}. For example, synchrotron emission typically has a steep falling spectrum\footnote{All spectral indices in this paper are in flux density units ($S \propto \nu^{\alpha}$).} ($\alpha\!\approx\! -0.5$ to $-1.0$), while free-free emission has a flatter spectrum when in the optically thin regime at gigahertz frequencies (see \S\ref{sec:fit_strategy}, for details).

\change{While synchrotron, free-free and thermal dust emission are well-understood at this point, discussions on the origin of spinning dust emission (or AME in general) are ongoing. The main carrier of this emission is thought to be polycyclic aromatic hydrocarbons (PAHs) due to their small size, significant dipole moments, and abundance within the ISM \citep{DraineLazarian1998, Dickinson2018, Hensley2021}. Yet several analyses have found little correlation between AME and PAH tracers \citep[eg.,][]{Tibbs2011,Vidal2011,Tibbs2012,Battistelli2015,Hoang2016}, which suggests that this may not be the case. However, this does not necessarily mean that PAHs are not the carriers of AME because the observed lack of correlation may be due to different excitation physics of PAHs in different environments \citep{Hensley2021}; there might be very large and currently under-appreciated differences in the emissivity of PAHs in different interstellar environments, for example.  Alternatives, such as nanosilicates have also been shown to be viable AME carriers \citep{Hensley2017} and nanodiamonds have also been reported to potentially carry spinning dust emission at least in some circumstellar environments \citet{Greaves2018}. Close comparisons between high angular resolution radio and infra-red data remains an important way to identify AME carriers.}

Large-scale, total-power Galactic radio surveys have up to now, for the most part, been conducted at frequencies of a few gigahertz or below  \citep{Haslam1982,ReichReich1986,Jonas1998,Calabretta2014,Carretti2019}. Even with the largest radio dishes, the angular resolution at these frequencies is \change{modest---typically} tens of arcmin or larger. The Canadian Galactic plane surveys at 408\,MHz and 1.4\,GHz have combined single-dish with interferometric data to achieve an angular resolution of 1\,arcmin \change{\citep{Landecker2010,Tung2017}}. Similarly, the GLOSTAR survey combines VLA and Effelsberg data at 4--8\,GHz to achieve sub-arcmin resolution \citep{Brunthaler2021}. At Ka-band (26--40\,GHz) near 30\,GHz, the only total power surveys available are the full-sky maps from WMAP \citep{Bennett2013} and \textit{Planck} \citep{Planck2018I} with angular resolutions of $\approx 32^\prime$. A 33\,GHz interferometric survey using the Very Small Array \citep{TodorovicEtAl2010} had an angular resolution of 13\,arcmin but only covered longitudes $26^{\circ}$--$46^{\circ}$.

The lack of high resolution radio data at high ($\gtrsim 10$\,GHz) frequencies has been partly due to the lack of sensitive receivers but also the difficulty of observing from the ground above a few gigahertz. The atmosphere becomes increasingly opaque due to strong absorption from water and oxygen in some bands---most notably, the 22\,GHz water line and 61\,GHz oxygen line. Although interferometric observations are possible at these high radio frequencies \citep[e.g., AMI at 12--18\,GHz;][]{Perrott2015}, mapping diffuse emission on large angular scales is difficult or impossible. Only total power imaging with single-dish telescopes can easily map the sky on larger angular scales. A focal plane array of detectors is therefore the ideal solution to mapping the Galaxy to higher frequencies. The CO-Mapping Array Project (COMAP) Pathfinder is such an instrument. 

\change{Focal plane arrays are ideal for mapping large areas, allowing multiple independent observations of the sky to take place at once. This is especially important for ground-based observations at high frequencies ($\gtrsim 10$\,GHz) due to increased noise levels from atmospheric contributions to the system temperature ($T_{sys}$), $1/f$ noise and smaller beamwidths.}

COMAP was primarily designed to map the highly redshifted CO emission for understanding star formation and galaxy evolution over cosmic time \citep[e.g.,][]{LiEtAl2015}. It uses a wide band covering 26--34\,GHz with a total of 4096 channels, which provides sensitivity to the 115\,GHz $J=1-0$ CO line at $z=3.4$--4.4. For Galactic observations, the large bandwidth and focal plane array of 19 detectors is ideal for mapping the large-scale 30\,GHz continuum as well as some spectral lines\change{, such} as radio recombination lines (RRLs). The 26--34\,GHz range is a particularly interesting choice for a Galactic survey. First, the Galaxy has never been surveyed at this frequency and angular resolution. Second, the relatively high frequency is ideal for quantifying star formation based on the level of free-free emission present \citep[e.g.,][]{MurphyEtAl2010}. At lower radio frequencies ($\sim 1\,$GHz and below) ultra-compact H\textsc{ii} (UCH\textsc{ii}) regions are optically thick and therefore go undetected, while at 30\,GHz the majority of sources will be in the optically thin regime \citep[e.g.,][]{Kurtz1994}. Third, anomalous microwave emission (AME), which is thought to be primarily due to spinning dust emission \citep[][]{DickinsonEtAl2018}, has a peaked spectrum near 30\,GHz. Therefore, the COMAP survey is an ideal tool to map the AME and study how it varies with interstellar environments. The complete survey, which will be made publicly available by 2023/24, will be a valuable resource for Galactic astronomers.

This paper is part of a series of papers from the COMAP collaboration \citep{es_I}, which outline the instrument and operations, as well as the cosmological results and interpretation. We envisage a series of future papers from the Galactic survey, covering specific sources and areas of investigation, culminating in the release of the complete COMAP Galactic plane survey covering $\ell \sim 20\degree$--$220\degree$.

This paper is organized as follows. \S\ref{sec:obs} gives an overview of the COMAP instrument and the observations, data processing, calibration, and map-making. \S\ref{sec:ancillary} summarises the various multi-frequency ancillary datasets that are used in conjunction with the COMAP data. \S\ref{sec:sed_fitting} describes the source extraction, photometry to produce SEDs for bright sources in the map, and SED model fitting. \S\ref{sec:results} presents the results including an \change{overview} of current COMAP maps (\S\ref{sec:results_comap}), correlations with other surveys (\S\ref{sec:results_correlation}), UCH\textsc{ii} analysis (\S\ref{sec:results_uchii}), SEDs of H\textsc{ii} regions (\S\ref{sec:results_hii}),  supernova remnants (\S\ref{sec:results_snr}), and RRLs (\S\ref{sec:results_rrls}). Finally, in  \S\ref{sec:discussion} we discuss the results so far and conclude with a future outlook.

\begin{table}
\caption{Main characteristics and parameters of the COMAP Galactic plane survey.}\label{table:characteristics}
\centering
 \begin{tabular}{lc} 
 \hline\hline
 Parameter & Value \\
 \hline
 Frequency Coverage & $26$--$34\,$GHz \\
Channel Bandwidth & 2\,MHz \\
Binned Bandwidth & 1\,GHz \\
System Temperature & 30--44\,K \\
Beam FWHM @ 30\,GHz  & $4.5^{\prime}$ \\
Beam Gain & $40$--70\,Jy K$^{-1}$ \\
Beam Efficiency &  $0.72 \pm 0.01$ \\
Absolute Calibration Accuracy & $3.2 \pm 0.1\,\%$ \\
Relative Calibration Accuracy & $<1\,\%$ \\
\hline
Galactic Longitude Range & $20^{\circ} < \ell < 220^\circ$ \\
Galactic Latitude Range & $-2^\circ < b < 2^{\circ}$ \\
 \hline
 \end{tabular}
\end{table}

\section{Instrument and Observations Overview} 
\label{sec:obs}

In this Section we will give a brief overview of the instrument, observations, and data processing relevant to the COMAP Galactic plane survey. The data processing of the Galactic plane survey differs in several regards to the cosmology survey as we wish to preserve the continuum emission. The primary differences are in respect to how time correlated noise fluctuations ($1/f$ noise) are suppressed using high-pass filters (\S\ref{subsec:dataOverview}) and destriping map-making (\S\ref{sec:mapmaking}), and the calibration of the data using astronomical sources (\S\ref{subsec:Calibration}). An overview of the instrument and survey parameters is given in Table~\ref{table:characteristics}.

The final COMAP Galactic plane survey maps have eight frequency bands at: 26.5, 27.5, 28.5, 29.5, 30.5, 31.5, 32.5, and 33.5\,GHz. When describing the maps in later sections we will refer to the 30.5\,GHz map for comparisons with other data, however all eight bands are used when fitting the spectral energy distributions of the sources discussed in \S\ref{sec:results}. \change{Note that throughout this paper we} use 30\,GHz when referring to the 30.5\,GHz COMAP data.

\subsection{Instrument}\label{sec:instrument}

Observations for the COMAP Galactic plane survey were made using the COMAP Pathfinder telescope \citep{es_II}, sited in Owens Valley Radio Observatory (OVRO) in California. The telescope itself is of Cassegrain design and has a full-width half-maximum (FWHM) beam width of $4.5^{\prime}$ \change{consistent to $\pm4$\,\% across} \mbox{$26$--$34$\,GHz} \change{\citep{es_II}}. The pathfinder has a focal plane array (FPA) of 19 forward-facing pixels in a hexagonal pattern. Feed centers are separated by $65$\,mm giving a sky-angle offset between pixels of $12.04^{\prime}$ allowing for 19 independent observations of the sky to be taken simultaneously.

The radio frequency (RF) signal from each feed is passed into a polarizer which converts the left-circular wave into the linear polarization accepted by a low-noise amplifier operating at \mbox{$15$--$18$\,K}. The outgoing noise-wave from the amplifier has its polarization reversed on reflection by the secondary and therefore does not couple back to the amplifier to cause a ripple in the spectrum. The amplifier output is then down-converted in two stages. The first mixes the \mbox{$26$--$34$\,GHz} with a \mbox{24\,GHz} local oscillator (LO), producing the first intermediate frequency (IF) signal at \mbox{$2$--$10$\,GHz}. This is split in two paths, one feeding a \mbox{$2$--$6$\,GHz} filter (band A), and the other a \mbox{$6$--$10$\,GHz} filter (band B). Band A is mixed with a \mbox{$4$\,GHz} LO to produce an in-phase (I) and quadrature (Q) signal at \mbox{$0$--$2$\,GHz}. Each of these baseband signals is sampled in an 8\,bit ADC and the I and Q signals combined in an FPGA (field programmable gate array) to produce the lower sideband (LSB) of the \mbox{$4$\,GHz} LO at \mbox{$2$--$4$\,GHz}, and the upper sideband (USB) at \mbox{$4$--$6$\,GHz}. Similarly, band B uses a \mbox{$8$\,GHz} LO to convert to I and Q baseband, converted by the FPGA to produce the \mbox{$6$--$8$\,GHz} and \mbox{$8$--$10$\,GHz} sidebands. The four signals LSB:A, USB:A, LSB:B, and USB:B, correspond to four contiguous \mbox{$2$\,GHz} bands between \mbox{$26$--$34$\,GHz} on the sky.

The FPGAs then perform the spectral analysis on 1024 channels for each $2$\,GHz band, with $2$\,MHz spectral resolution across the full $8$\,GHz bandwidth observable with COMAP. For a more complete \change{description see} \cite{es_II}. The resulting dataset is collated alongside \change{pointing, environmental and housekeeping} data and stored locally at Caltech. 

\subsection{Observations} \label{subsec:observations}

The COMAP Galactic plane survey will \change{cover} the Galactic plane over the range $20^\circ < \ell < 220^\circ$ in longitude and $-2^\circ < b < 2^\circ$ in latitude. Observations started in June 2019 and are ongoing. Typically 1--2\,hours per day of COMAP observing time is dedicated to the Galactic plane survey. Since June 2019 we have surveyed between $20^\circ < \ell < 50^\circ$ in Galactic longitude, totalling $834$\,hours observing time. Observations are initially calibrated at the beginning and end of each observation using a thermal load with a known temperature \citep[details in][]{es_III} and then calibrated to an astronomical brightness scale using daily observations of the supernova remnant Taurus\,A/Crab nebula (Tau\,A, \S\ref{subsec:Calibration}). All observations were taken during the day.

The survey was conducted by observing the Galactic plane in discrete patches, where each patch covers an area of approximately 4\,deg$^2$. As the COMAP instrument focal plane spans approximately $\sim1^\circ$, we nested neighbouring patches to ensure a uniform sensitivity across the survey. To map out each patch we would begin scanning \change{the telescope} in the horizon frame with a Lissajous pattern, and allow the natural rotation of the sky to move the patch center through the field-of-view of the telescope. The Lissajous scanning strategy traces sinusoidal patterns in azimuth and elevation that result in each pixel in the celestial frame being visited by many different scan paths, a condition that is required for the map-making method (\S\ref{sec:mapmaking}) as it allows for the true sky signal and correlated noise to be separated. 

The Lissajous pattern is defined by the pair of equations

\begin{equation}\label{eq:LissAz}\delta_{A} = \frac{R}{\cos(E)}\sin(\omega_A t + \phi),\end{equation}
\begin{equation}\delta_{E} = R\sin(\omega_E t),\end{equation}
where $\delta_{A}$ and $\delta_{E}$ represent the offset in azimuth and elevation from the central azimuth ($A$) and elevation ($E$) coordinates, $\omega_E$ and $\omega_A$ are the angular \change{velocities along} each axis, $R$ is the radius of the scans, and $\phi$ is the relative phase. We used $R = 0.8^\circ$ for the entire survey, the phase was alternated between $\pm \frac{\pi}{2}$, and the ratio of the angular velocities was randomly selected in the range $0.6 < \omega_E/\omega_A < 1$. The telescope was driven close to its maximum rate in elevation ($V_E^\mathrm{max} = 0.5^\circ\,\mathrm{s}^{-1}$) for all observations. We modulated the Lissajous pattern by changing the azimuth scan speed ($V_A^\mathrm{max} = 1^\circ\,\mathrm{s}^{-1}$). 

\subsection{Data Processing}\label{subsec:dataOverview}

The nominal on-sky system temperatures were measured to be in the range $30$--$44$\,K across the full $26$--$34$\,GHz COMAP band. We rejected a number of channels that had abnormally high system temperatures  caused either by spectral aliasing at the edges of the bands or by resonances within the feed optics \citep[see][]{es_II}. After flagging bad channels, we averaged the native 2\,MHz channels into wide 1\,GHz bands.

When using the COMAP system for continuum science, the data are contaminated by substantial time-correlated noise (often referred to as $1/f$ noise; \citealt{Harper2018}) from two sources: fluctuations in the precipitable water vapor content in the atmosphere above the telescope, and small fluctuations in the gain or system temperature of the receiver low-noise amplifiers. Mitigating $1/f$ noise is critical to recovering the large-scale, diffuse Galactic structures and for achieving the lowest possible noise levels in the map. The strongest atmospheric $1/f$ noise can be mitigated by discarding  data from days that have either poor or turbulent weather conditions. These can be determined by measuring the feed-to-feed noise correlation (as the near-field atmospheric fluctuations will be strongly correlated between feeds), tracking the optical depth of the atmosphere using sky-dips \citep{Rohlfs2000}, and measuring the power spectrum of the data to determine $1/f$ noise properties. We identified the very worst observations where the atmospheric fluctuations were several times the white noise of the receiver at timescales of 10\,seconds or more and removed \change{them, cutting the TOD for all 8 channels. This amounted to cutting} approximately 17\,\% of the observations\change{, which were found to contain} severe atmospheric contamination.

There is some contamination of the data from ground emission due to the local mountain ranges that lie to the east and west of OVRO. We characterised the ground emission profiles by performing 360$^\circ$ azimuth sweeps at fixed elevations. From these observations we could determine that the scale size of the ground emission is $\gg 1^{\circ}$, much larger than typical patches observed. On the scale of a single scan \change{\citep[the period between two repointings, i.e. $\approx 1\,^{\circ}$,][]{es_III}} we approximate the ground emission (and other azimuth correlated systematics) using a linear slope in azimuth, which gives a median amplitude for the ground emission across the survey of $\approx 6$\,mK. \change{This ground emission slope is then subtracted from the TOD.}  

To suppress any remaining large-timescale $1/f$ noise fluctuations we use a running median filter with a scale size of 100\,seconds (approximately equal to the typical time needed to complete a full scan---\S\ref{subsec:observations}). Finally, we use the destriping map-making technique (see \S\ref{sec:mapmaking}) in combination with \change{the} observing strategy to suppress $1/f$ noise down to timescales of $1$\,second.

\subsection{Calibration}\label{subsec:Calibration}

Calibration of the COMAP Galactic plane survey is done in two steps. First we calibrate the data using the calibration vane, a remotely controlled microwave absorber acting as a thermal load that covers the feed array at the beginning and end of each observation (i.e., approximately every 40\,minutes). We take the difference between the known vane temperature ($T \approx 290$\,K) and the cold sky ($T \approx 2.7$\,K) to estimate both the system temperature and the total gain of the receiver and antenna system. The vane calibration procedure is described in \citet{es_III}.

The effect of atmospheric absorption at 30\,GHz is significant. We track atmospheric opacities using sky dips, which involves slewing \change{the} telescope between elevations $40\degree$ and $60\degree$ over a period of a few seconds. We find that the typical opacity of the atmosphere at OVRO is in the range $\tau \sim 0.07$--0.09, which equates to 15\,per\,cent absorption at $30\degree$ elevation. The vane calibration procedure corrects for atmospheric absorption due to the atmosphere along the line of sight. After the vane calibration we estimate the residual optical depth using the vane-calibrated observations of Taurus\,A/Crab nebula (Tau\,A) and Cassiopeia\,A (Cas\,A), and find the residual effect of atmospheric absorption to be $\delta\tau \sim 0.02 \pm 0.01$, which equates to a 2--3$\%$ residual uncertainty in the calibration at the elevations used for the COMAP Galactic plane survey.

Absolute calibration to the main beam brightness scale is done using Tau\,A. We observe Tau\,A once per day and fit the peak brightness using a 2-D Gaussian model. We derive the absolute calibration from the Tau\,A measurements by comparing with the WMAP spectral fits and secular decrease models of Tau\,A \citep[tables 16 and 17 of ][]{Weiland2011}. \change{We do not apply color corrections when calculating model fluxes to calibrate on, as these are $\ll1\,\%$ in each 1\,GHz band.} We verified the absolute calibration using Cas\,A and Jupiter and find an overall accuracy of $3.2\,\%$ with a relative calibration between bands of $<1\,\%$. Models for Cas\,A were also taken from \citet{Weiland2011}. The flux density model of Jupiter was derived by fitting a power-law to the  brightness temperature measurements taken using the CARMA instrument between 27 and 33\,GHz \citep{Karim2018}, and using the ephemeris of Jupiter to track its solid angle on the sky. \change{We combine these errors in quadrature giving an approximate 5$\%$ error on each of the 8 maps in this work.}

The COMAP beam has very good main beam isolation with the first sidelobe at more than 20\,dB below the main beam response. Although this may be good enough in terms of confusion of nearby bright sources, these sidelobes still result in a scale-dependent calibration \citep[see e.g.,][for a discussion of this issue]{Du2016}. To estimate the level of the effect we radially integrated the COMAP beam models described in \citet{es_II} out to the first null (i.e., the main beam) and third sidelobe (approximately $30^{\prime}$ from the line of sight). We found that the integrated power changes by $10\,\%$ between main beam and $30\,^\prime$ ($\approx 6\times$ the beam FWHM). For these early results we are largely interested in sources that are either unresolved or only partially resolved by the main beam, so the effect of beam dependence in this work is minimal. 

\subsection{Map Making} \label{sec:mapmaking}

In order to suppress spurious large-scale contamination and any remaining $1/f$ noise within the COMAP continuum data we have implemented a bespoke destriping map-maker using the methods outlined in \citet{Delabrouille1998}, \citet{Sutton2009}, and \citet{Sutton2010}. Destriping map-making solves for $1/f$ noise by fitting linear offsets to the time-ordered data, but uses the scanning information to separate the $1/f$ noise from the true sky signal. To illustrate how destriping map-making works we will define the following data model

\begin{equation}\label{eqn:datamodel} 
  \mathbf{d} = \mathbf{P} \mathbf{m} + \mathbf{F} \mathbf{a} + \mathbf{n}_w,
\end{equation}
where $\mathbf{d}$ is a vector containing the time-ordered data \change{(for a single band, i.e. one time-stream)}, $\mathbf{m}$ is the true sky signal, $\mathbf{P}$ maps the sky signal to the time domain, $\mathbf{a}$ are the offsets that describe the $1/f$ noise and $\mathbf{F}$ maps these offsets to $\mathbf{d}$, and finally $\mathbf{n}_w$ is the uncorrelated white noise vector.

Solving for the offsets in Equation~\ref{eqn:datamodel} results in

\begin{equation}
  \hat{\mathbf{a}} = \left(\mathbf{F}^T\mathbf{Z}^T \mathbf{N}^{-1} \mathbf{Z} \mathbf{F}\right)^{-1} \mathbf{F}^T \mathbf{Z}^T \mathbf{N}^{-1} \mathbf{Z} \mathbf{d}, 
\end{equation}
where $\hat{\mathbf{a}}$ is the maximum-likelihood estimate of the offsets, and $\mathbf{N} = \left<\mathbf{n}_w \mathbf{n}_w^T\right>$ is the white noise covariance matrix. The matrix $\mathbf{Z}$ is defined as 

\change{\begin{equation}
  \mathbf{Z} = \mathbf{I} - \mathbf{P}\left(\mathbf{P}^T \mathbf{N}^{-1} \mathbf{P}\right)^{-1} \mathbf{P}^T \mathbf{N}^{-1}.
\end{equation}}

The COMAP Galactic plane survey maps use a plane cylindrical polar projection in the Galactic coordinate frame. The nominal offset length used to destripe the data was 1\,second, which corresponds to approximately $18^{\prime}$ scales on the sky. Our destriping map-maker suppresses the $1/f$ noise by a factor of four or better on scales up to $30^{\prime}$. The average noise level of the final maps is $4.6\pm1.2$\,mK\,arcmin$^{2}$.

\section{Ancillary Data} 
\label{sec:ancillary}
Ancillary data were used from a variety of different single-dish experiments between $408$\,MHz and $25$\,THz ($\lambda=12\,\mu$m). Spectral models were only fitted to flux densities below $3$\,THz ($\lambda=100\,\mu$m) as higher frequencies are contaminated by \change{stochastic heating and emission from Polycyclic Aromatic Hydrocarbons (PAH)}, which causes higher fluxes than would be naively expected from the simple models described in \S\ref{sec:sed_fitting}. In addition some frequencies were not fitted due to contamination from spectral lines, such as the \textit{Planck} 100 and 217\,GHz channels (contaminated by Galactic CO lines). A summary of the datasets used in this work is given in Table~\ref{tab:surveys}.

\subsection{Map preparation} \label{subsec:mapPrep}

The map-space processing was done using a custom designed python pipeline, which operates in two steps: preparing and smoothing the maps, and then performing source extraction and photometry. This process began with any map-specific processes (such as reprojection or beam-correction) and was then followed by two common final preparation steps in all maps.

In order to account for different beam sizes the data were smoothed to a common resolution---in this analysis, $5^{\prime}$. The units of each map were also converted to a common unit (MJy\,sr$^{-1}$) to allow easy extraction of integrated flux densities, and for maps to be readily compared side-by-side.

\subsection{Ancillary Maps}\label{sec:ancmap}

\subsubsection{Low-frequency datasets}

Beginning at low frequencies, we used the Effelsberg Galactic plane survey  $2.7$\,GHz ($\lambda=11$\,cm) maps, with associated $10$\,\% calibration error \citep{ReichEtAl1984,ReichEtAl1990,FurstEtAl1990}. This calibration was based on three point sources; two Seyfert 1 galaxies (3C\,286 and 3C\,138) and the quasar 3C\,48, and large-scale gradients were removed by comparing with the Stockert survey \citep{ReichReich1986}.

We also used the Parkes $5$\,GHz Galactic plane survey \citep{HaynesEtAl1978} to constrain the free-free contributions to the overall source SEDs. For calibration we took the stated $8$\,\% accuracy based on measurements of Hydra\,A (assumed to be $13$\,Jy). In order to further verify this and ensure a reasonable calibration on scales greater than those of the beam \change{(to} be able to judge source morphology etc.) we smoothed \change{the} Parkes and the Sino-German  Survey \citep{GaoEtAl2019} to $10\,^\prime$ resolution and performed T-T plot analysis \citep[e.g.,][]{Davies1996} \change{on the Galactic plane between $20<\ell<40\,^\circ$ for galactic latitudes less than $|b|<2\,^\circ$}, \change{masking out bright point sources, and} fitting for a gradient between the maps of $(0.98\pm0.03)$ consistent with a one-to-one ratio. \change{Since both surveys are comparable single-dish $5$\,GHz surveys, a gradient consistent with $1$ confirms both that Parkes is correctly calibrated, and that the diffuse structures seen in the Sino-German Survey are present in Parkes $5$\,GHz data.}

For $10$\,GHz data points we incorporated the Nobeyama FUGIN Galactic plane survey \citep{HandaEtAl1987}. The survey was ideally placed for this analysis with a slightly smaller beamwidth of $2.7^{\prime}$ than COMAP on a single-dish telescope. No formal calibration uncertainties were given in \cite{HandaEtAl1987}; we adopt a 10\% calibration uncertainty, which results in reasonable $\chi^2$ values for the fitted SED models (\S\,\ref{sec:results_hii}).

It may be noted that the Canadian Galactic Plane Survey \citep[CGPS;][]{TaylorEtAl2003} at $1.4$\,GHz was not included in our analysis despite having arcminute resolution. This was because the survey does not cover the longitude range analyzed in this work, but it will be  valuable for future analyses at higher longitudes $(\ell > 58^{\circ}$).

\subsection{High-frequency maps}
Including the \textit{Planck} HFI bands to fill in the thermal dust emission was vital to understanding the sources presented in this work. Full-mission \change{maps were} used for the analysis presented, and were sourced from the \textit{Planck} Legacy Archive.~$\!\!\!\!$ \footnote{Website: https://pla.esac.esa.int/\#home} Maps from \textit{Planck} HFI bands were convolved to a $5^\prime$ gaussian beam using the HFI beam models \citep[see][for details]{Planck2013VII} and then reprojected from HEALPix \citep{GorskiEtAl2005} to a Cartesian grid. Due to the high-resolution of the HEALPix maps retrieved and our later need to further smooth the maps we do not see significant pixelisation effects when comparing \textit{Planck} data against other ancillary datasets. We also used color correction as described by \citet{Planck2013IX} for HFI which is accounted for in the MCMC fitting. Furthermore we used calibration errors as provided in \citet{Planck2013IX}, which give calibration errors between $0.09$ and $6.4$\,\% for the HFI. 

Due to the beamwidth of the \textit{Planck} instrument being significantly larger at lower frequencies, we were only able to employ the HFI bands above $217$\,GHz for this analysis. This unfortunately meant that we had no surveys between the upper COMAP band ($33.5$\,GHz) and the $217$\,GHz \textit{Planck} band to aid fitting the AME component.% For this purpose, a further $1^\circ$ analysis is performed on the W43 region as a whole incorporating the lower two HFI bands ($100$ and $143$\,GHz) and the LFI $30$, $44$ and $70$\,GHz maps. The LFI bands are processed in a similar way to the HFI, but with the color corrections and calibration factors provided by \citet{Planck2013V}.

While we include all the data points in the SED plots, we did not use the \change{\textit{Planck} $217$\,GHz band} in fitting since these bands are affected by Galactic CO(1--0) and CO(2--1) transitions. For future work we will attempt to include these frequencies by subtracting the \textit{Planck} internal estimate of Galactic CO emission \citep{Planck2013XIII}.

Finally, we utilized \textit{Akari} \citep{DoiEtAl2015} and IRIS \citep{Miville-DeschenesLagache2005} data to fill in terahertz frequencies. These maps were sourced via the CADE archive, retrieved in HEALPix and then converted to a cartesian grid. As seen in Table\,\ref{tab:surveys} we did not use all the terahertz maps for SED fitting due to stochastic heating within the dust cloud taking effect at frequencies higher than that of the modified blackbody peak. For this purpose, we found that placing a limit at $3000$\,GHz allows for enough data points to be included such that the modified blackbody was sufficiently constrained but that we did not include the more complex reemitted radiation \citep[see in-depth discussion in][]{CompiegneEtAl2011}. For the two surveys we take calibration errors from their respective papers as reported in Table \ref{tab:surveys} giving values of between $5.7$\,\% and $8.9$\,\% for \textit{Akari} and $13.5$\,\% for IRAS band 4.

\begin{table*}
\setlength{\tabcolsep}{3pt}
 \caption{Datasets used for SED extraction, alongside their sky coverage, assumed calibration error and native resolutions.} %Surveys marked with a star ($\star$) were extracted for reference, but not used in fitting due to contamination by strong line emission or thermal reabsorption.}
 \label{tab:surveys}
 \begin{tabular*}{\textwidth}{lccccl}
  \hline
  Survey & Frequency & Calibration Error & Coverage & Resolution & References \\
  & [GHz] & [\%] & & [$^\prime$] \\
  \hline
  Effelsberg    \dotfill & $2.7$ & $10$ & $357<\ell<240\,^\circ$, $|b|<5\,^\circ$ & $4.3$ & \citet{ReichEtAl1984,ReichEtAl1990,FurstEtAl1990} \\
  Parkes    \dotfill & $5.0$ & \change{$8$} & $190<\ell<40^\circ$, $|b|<2^\circ$ & $4.3$ & \citet{HaynesEtAl1978} \\
  Nobeyama      \dotfill & $10.0$ & $10$ & $10<\ell<50^\circ$, $|b|<1^\circ$ & $2.7$ & \citet{HandaEtAl1987} \\
  COMAP         \dotfill & $26.5$ & $5$ & $20<\ell<40^\circ$, $|b|<2^\circ$ & $4.5$ & \textit{This work} \\
  COMAP         \dotfill & $27.5$ & $5$ & $20<\ell<40^\circ$, $|b|<2^\circ$ &  $4.5$ & \textit{This work} \\
  COMAP         \dotfill & $28.5$ & $5$ & $20<\ell<40^\circ$, $|b|<2^\circ$ &  $4.5$ & \textit{This work} \\
  COMAP         \dotfill & $29.5$ & $5$ & $20<\ell<40^\circ$, $|b|<2^\circ$ &  $4.5$ & \textit{This work} \\
  COMAP         \dotfill & $30.5$ & $5$ & $20<\ell<40^\circ$, $|b|<2^\circ$ &  $4.5$ & \textit{This work} \\
  COMAP         \dotfill & $31.5$ & $5$ & $20<\ell<40^\circ$, $|b|<2^\circ$ &  $4.5$ & \textit{This work} \\
  COMAP         \dotfill & $32.5$ & $5$ & $20<\ell<40^\circ$, $|b|<2^\circ$ &  $4.5$ & \textit{This work} \\
  COMAP         \dotfill & $33.5$ & $5$ & $20<\ell<40^\circ$, $|b|<2^\circ$ &  $4.5$ & \textit{This work} \\
  \textit{Planck} HFI (DR3.1) \dotfill & $353$ & $0.78$ & all sky & $4.8$ & \citet{Planck2018I} \\
  \textit{Planck} HFI (DR3.1) \dotfill & $545$ & $6.1$ & all sky & $4.7$ & \citet{Planck2018I} \\
  \textit{Planck} HFI (DR3.1) \dotfill & $857$ & $6.2$ & all sky & $4.3$ & \citet{Planck2018I} \\
  \textit{Akari} ($160\,\mu$m) \dotfill & $1875$ & $8.9$ & all sky & $1.5$ & \citet{DoiEtAl2015,TakitaEtAl2015} \\
  \textit{Akari} ($140\,\mu$m) \dotfill & $2143$ & $8.9$ & all sky & $1.5$ & \citet{DoiEtAl2015,TakitaEtAl2015} \\
  IRAS (IRIS) Band 4 ($12\,\mu$m) \dotfill & $3000$ & $13.5$ & all sky & $4.3$ & \citet{Miville-DeschenesLagache2005} \\
  \textit{Akari} ($90\,\mu$m) \dotfill & $3333$ & $5.7$ & all sky & $1.0$ & \citet{DoiEtAl2015,TakitaEtAl2015} \\
  \hline
 \end{tabular*}
\end{table*}
\setlength{\tabcolsep}{6pt}

\section{Source Selection and SED Model Fitting}
\label{sec:sed_fitting}

For our early analysis, we perform source extraction and SED fitting to a number of H\textsc{ii} regions and SNR. For H\textsc{ii} regions we utilise a source extraction and aperture photometry pipeline as described in \S\ref{subsec:SE} and \S\ref{sec:aperturephot}. For SNR we fit apertures and annuli by eye and then run them through the aforementioned aperture photometry pipeline.

\subsection{Source extraction} \label{subsec:SE}

The first task required identifying bright, relatively compact, sources within a small sample region of the COMAP survey for early analysis. \change{We} focus on sources that can be measured easily due to their brightness relative to their background, and compact size relative to that of the beam, i.e. $\lesssim 5^{\prime}$. To do this the COMAP $30$\,GHz map was put through the preparation described in \ref{subsec:mapPrep} and then put through a python implementation of Source Extractor Python \citep[SEP;][]{Barbary2016}, a python wrapper for SExtractor \citep{BertinArnouts1996}.

Within the SEP implementation several options were invoked to ensure a reliable source list was returned which minimized the number of false detections. The first process used was filtering---this was done with an un-normalized Ricker wavelet of the form

\begin{equation} f_{MH}(r, \sigma) = \left(1-\frac{1}{2}\frac{r^2}{\sigma^2}\right)\exp\left\{-\frac{1}{2}\frac{r^2}{\sigma^2}\right\}, \end{equation}
where $r$ represents the radius from the center of the filter, and $\sigma$ represents the filter's standard deviation. The filter was computed on a $15\times15$ array (which equates to $15^\prime \times15^\prime$ space when projected onto the map) with a standard deviation of $2.2^\prime$ to match the COMAP beam.

When inputting a signal-to-noise cutoff into SEP, we find that the influence of \change{the} diffuse background and the high density of sources causes the noise estimates to be greater than those measured by \change{using an} aperture on the background by an order of magnitude. As such, we input a measured background noise level of $6.5\pm2.4$\,mK on scales of $15\,^\prime$ and applied a $5\sigma$ detection limit on all sources.

%The maps were then deblended - which we found works best when increasing the threshold to very high values, and keeping the contrast ratio around $1$\,\%. The final extraction was found to be best when using a threshold of $12000$ and a contrast value of $0.01$ - which ultimately was able to pick out sources above the background diffuse emission most efficiently. The maps were also cleaned by SEP with a cleaning value of $0.5$.

After this analysis of the raw COMAP $30$\,GHz map 21 candidate sources were identified. Through checking each source by eye in all 8 COMAP bands and the $5$\,GHz Parkes and \textit{Planck} maps, sources were either verified or rejected. These sources were verified by a clear detection in all 8 COMAP bands in addition to a clear detection in one the two other maps (verifying a strong dust and/or free-free component from the source). Nevertheless, performing photometry on some of these remaining sources was still non-trivial. We therefore further removed additional sources that were still weak and/or if the background was sufficiently bright/complicated that the flux densities were not robust to variation in the background annuli locations. This left a total of nine sources with robust SEDs to present in \S\ref{sec:results_hii}.

\subsection{Aperture Photometry}\label{sec:aperturephot}

We use aperture photometry to measure the flux densities of the sources identified in the previous section. While we see that Gaussian fitting may be more suited to such a complex region as the Galactic plane, particularly for sources embedded within extended emission, we implement a simple aperture photometry technique here to provide an early indication of the spectral composition of such sources on small scales. Since we are focused on brighter sources, aperture photometry should provide a reliable estimate of the flux density.

For the H\textsc{ii} regions, we use a constant aperture radius of $r=5^{\prime}\,(1\sigma)$ to integrate the flux of each source. We find this to be suitable given that each source on the initial extracted list has a fitted standard deviation in the range $2.2^\prime<\sigma<2.5^\prime$, giving a $\approx2\sigma$ aperture radius size. For background measurements we use an annulus, drawn between $6.7^\prime$ and $8.3^\prime$ from the source coordinates and use the median pixel value to estimate the background to be robust against nearby bright sources.

In the case of SNR, we treat these separately due to the large amount of variability between sources, and the tendency for the SNR to appear as extended sources. For these sources we did not perform a specific source extraction, but used the values given in \citet{Green2019} to set initial values for the radius and central coordinates of the aperture and annulus used, adjusting these to fit the SNR as they appear in the maps and the background. We then used the adapted central coordinates and radii in the aperture photometry pipeline in \S\ref{sec:aperturephot} to obtain flux measurements from our maps.

\subsection{SED Fitting} 
\label{sec:fit_strategy}

We fit the flux densities of each H\textsc{ii} source measured using aperture photometry (\S\,\ref{sec:aperturephot}) with a three component 
model of the free-free ($S_\mathrm{ff}$), thermal dust ($S_\mathrm{td}$), and AME components ($S_\mathrm{AME}$):
\begin{equation}
S = S_\mathrm{ff} + S_\mathrm{td} + S_\mathrm{AME}.
\end{equation}

We model the free-free emission flux density by first calculating the free-free brightness temperature and converting to flux density units via
\begin{equation}
S_\mathrm{ff} = \frac{2k_B T_\mathrm{ff}\Omega_\mathrm{beam}\nu^2}{c^2},
\end{equation}
where $k_B$ represents the Boltzmann constant, $\Omega_\mathrm{beam}$ the beam solid angle, $\nu$ the frequency, and $c$ the speed of light. $T_\mathrm{ff}$ is defined as \citep[e.g.,][]{Draine2011}
\begin{equation}
T_\mathrm{ff} = T_e(1-e^{-\tau_\mathrm{ff}}),
\end{equation}
where $T_e$ is the the electron temperature and $\tau_\mathrm{ff}$ is the free-free emission optical depth
\begin{equation}
\tau_\mathrm{ff} = 5.468\times10^{-2}T_e^{-1.5}\nu^{-2}\mathrm{EM} g_\mathrm{ff}~, 
\end{equation}
where $g_\mathrm{ff}$ is the Gaunt factor, which for these frequencies can be approximated as 
\begin{equation}\label{eqn:gaunt}
g_\mathrm{ff} = \ln\left(\exp\left[5.960-\frac{\sqrt{3}}{\pi}\ln(\nu_\mathrm{GHz}T_4^{-1.5})\right] + 2.71828\right),
\end{equation}
where $T_4 = T_e \times 10^{-4}$. For the SED fitting we let the emission measure be a free parameter, while the electron temperature we fix to a typical value of $T_e = 7500$\,K \citep[e.g.,][]{Paladini2003}. However, the fit is only weakly dependent on the choice of $T_e$.

The thermal dust emission is modeled as a modified blackbody curve given by
\begin{equation}
S_\mathrm{td} = 2h\frac{\nu^3}{c^2}\frac{1}{e^{h\nu/k_B T_\mathrm{d}}-1}\left(\frac{\nu}{353\,\mathrm{GHz}}\right)^{\beta_\mathrm{d}}\tau_{353}\Omega_\mathrm{beam}~,
\end{equation}
and we fit for the optical depth of thermal emission at $353$\,GHz ($\tau_{353}$), the dust temperature ($T_\mathrm{d}$), and the emissivity spectral index ($\beta_\mathrm{d}$). \change{It is worth noting that there are well-known degeneracies between emissivity spectral index and dust temperature \citep[e.g.][]{Dupac2003, Desert2008, Planck2013XI}.}

We model the AME spectrum as a log-normal curve as this is a flexible model that is a good approximation to spinning dust spectrum \citep{Bonaldi2007,Stevenson2014,DickinsonEtAl2018}; it is defined as

\begin{equation}
S_\mathrm{AME} = A_\mathrm{AME}\exp\left\{ -\frac{1}{2}\left(\frac{\ln(\nu) - \ln(\nu_\mathrm{AME})}{w_\mathrm{AME}}\right)^2 \right\}~, 
\end{equation}
where we fit for the amplitude of the AME ($A_\mathrm{AME}$), the peak frequency ($\nu_\mathrm{AME}$), and the width of the log-normal ($w_\mathrm{AME}$).

For the \change{supernova remnants} we do not fit for a free-free or AME component but instead we fit for synchrotron emission ($S_\mathrm{sync}$) so the SED model becomes 

\begin{equation}
    S = S_\mathrm{sync} + S_\mathrm{td},
\end{equation}
where the synchrotron component is modeled as a simple power-law

\begin{equation}
    S_\mathrm{sync} = A_\mathrm{sync} \nu^\alpha,
\end{equation}
where we fit for the spectral index of the synchrotron ($\alpha$), and the synchrotron amplitude at 1\,GHz ($A_\mathrm{sync}$).

The SEDs were fitted using a Markov chain Monte-Carlo (MCMC) analysis, and the \texttt{emcee} ensemble sampler \citep{Goodman2010,Foreman-Mackey2013} as the backend. The implementation we use is the same as that described in \citet{Cepeda-ArroitaEtAl2021} with a modification to account for the correlation in the noise between COMAP bands. For each fit we use 300 chains, each with 5000 steps. Burn-in times for each fit required between 1300--1400 steps as assessed by eye, and so we discarded the first 1500 from each chain. The typical correlation length of the samples was 390 steps, which we used to thin the chains and to suppress sample-to-sample correlations. We checked each chain for convergence, and discarded those that failed. 

At the resolution of COMAP there are no surveys that cover the frequency range \change{40}--100\,GHz. The lack of data at these frequencies means that the peak and width of the AME log-normal are poorly constrained. To improve the convergence of the MCMC fitter we implement Gaussian priors on $\nu_\mathrm{AME}$ and $w_\mathrm{AME}$. The Gaussian priors are informed by measuring the integrated flux density of W43 \change{using the same $1\,^\circ$ radius aperture and background annulus ($1.3$--$1.7\,^\circ$) as used in \citet{Irfan2015} and \citet{Genova_SantosEtAl2017}. By using maps smoothed to $30\,^\prime$ resolution we were able to use the datasets in Table~\ref{tab:surveys} alongside the lower resolution \textit{Planck} LFI 28.4, 44, and 70\,GHz bands. Fitting for a model with free-free, AME and thermal dust emission  we constrain a mean peak frequency and log-normal width for the region.} The final Gaussian priors we used for the H\textsc{ii} SED fits were $N(28 \pm 15\mathrm{GHz})$ for the peak frequency ($\nu_\mathrm{AME}$), and $N(0.6 \pm 0.2)$ for the width ($w_\mathrm{AME}$).

\section{Results} 
\label{sec:results}

\subsection{COMAP Survey Map}
\label{sec:results_comap}

\begin{figure*}[ht]
\begin{center}
\includegraphics[width=\textwidth]{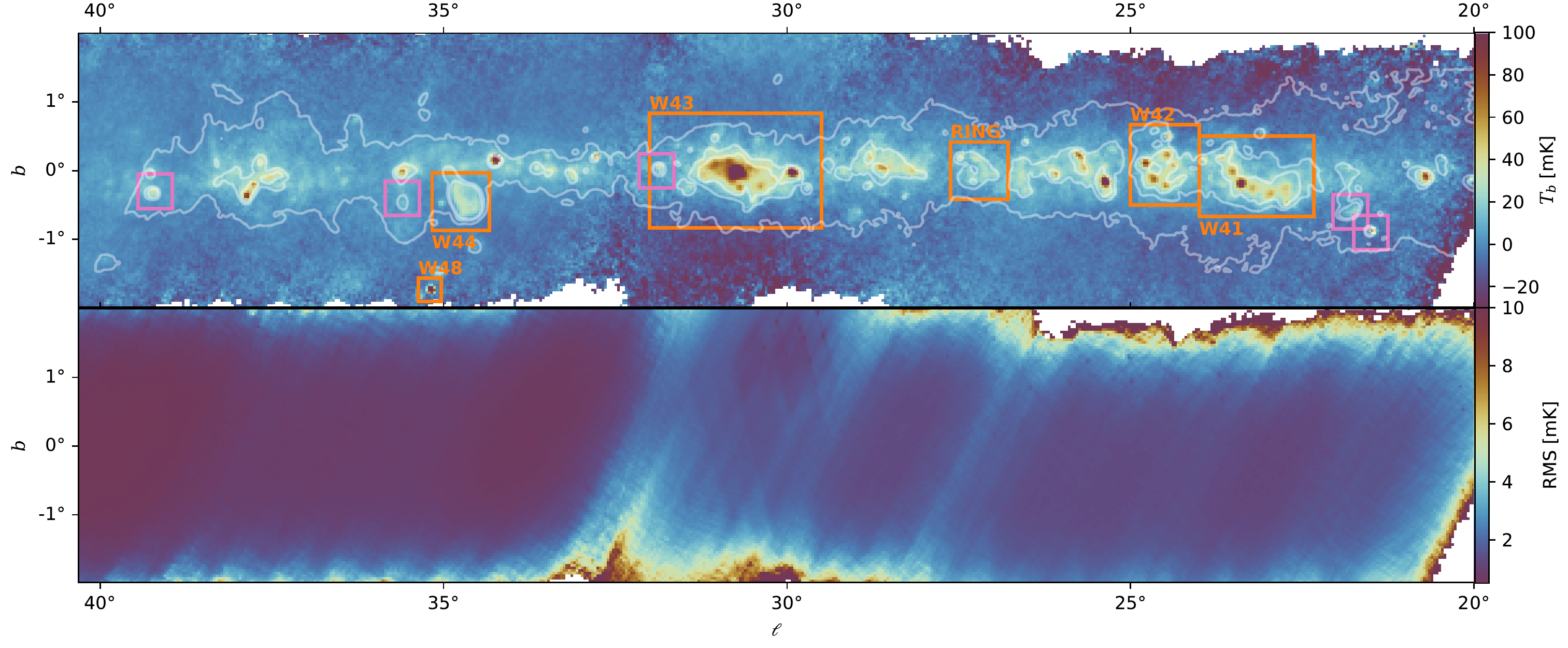}
\caption{The current COMAP band-averaged $30$\,GHz map (\textit{top} panel) covering the Galactic plane between $20\degree<\ell<40\degree$. The color scale is linear and in units of brightness temperature (mK). Masked pixels are white. The \textit{bottom} panel shows the coverage in terms of local RMS in the map. Westerhout complexes are indicated as \textit{orange rectangles}, including the SNR W44. The other detected SNRs are indicated by \textit{purple squares}. Contours are given for the Parkes $5$\,GHz Galactic plane survey at levels of $1$, $1.5$ and $2.0$\,MJy/sr.}
\label{fig:COMAP-hits+map}
\end{center}
\end{figure*}

In Fig.~\ref{fig:COMAP-hits+map} we show the COMAP 30.5\,GHz map covering $20\degree<\ell<40\degree$ on the \textit{top} panel, and the total number of integrations per 1$^{\prime}$ pixel in the \textit{bottom} panel. The map units are in mK and it has a resolution of $4.5^{\prime}$. We detect with high signal-to-noise ratio several bright giant molecular cloud regions such as G023.3$-$00.3 \citep[e.g.,][]{Messineo2014} and W43 at $(\ell,b) = (25.\!\!^{\circ}4,-0.\!\!^{\circ}2$) \citep[e.g.,][]{Nguyen2011}; several giant H\textsc{ii} regions such as G24.5$-$0.0 and W42 at $(\ell,b) = (25.\!\!^{\circ}4,-0.\!\!^{\circ}2)$, as well as many known H\textsc{ii} regions and H\textsc{ii} complexes \citep[e.g.,][]{Paladini2003,AndersonEtAl2014}; and also several supernova remnants (\S\ref{sec:results_snr}). As well as discrete sources we also detect extended emission in the form of diffuse structures and spurs. Most of the diffuse emission is caused by diffuse ionized gas formed from the leakage of ionizing radiation from nearby O/B stars into the ISM \citep{Zurita2000}, but there will also be  dust-correlated anomalous microwave emission (AME) due to spinning dust grains \change{\citep{DickinsonEtAl2018,Hensley2016}} potentially contributing up to $\approx$20--45\,\% of the observed brightness \citep{PlanckEarlyXXI,PlanckIntXXIII}. 

The map has a high signal-to-noise ratio (S/N) across much of the Galactic plane. The noise level of the map varies between \mbox{2--3\,mK\,beam$^{-1}$} away from the edges, or equivalently 0.1--0.15\,Jy\,beam$^{-1}$. Due to filtering of the time-ordered data on timescales corresponding to the typical scan length of $\sim \!1^{\circ}$, the largest scales are not well constrained. We find that there is a loss of flux of $15\%$ or more on scales larger than about $30^{\prime}$. For example, a comparison between the integrated flux density of W43 \change{at $30^{\prime}$ resolution} over a $2^{\circ}$ diameter aperture in the COMAP 28.5\,GHz data to that measured by \textit{Planck} LFI at 28.4\,GHz \citep[$504\pm26$\,Jy;][]{Irfan2015} shows that we are underestimating the integrated flux density by $\approx 15$\,\%. However, as discussed in \S\ref{subsec:Calibration} the data have been calibrated to 5\,\% or better at the main beam scale \change{(taking into account both opacity errors and those on the raw calibration)} out to a few times the FWHM ($\sim\!12^{\prime}$), therefore we will primarily focus on discrete \change{(i.e. clear, high S/N)} sources for this first analysis. In future releases we will use the COMAP beam models \citep{es_II} to deconvolve the map and use simulations to help improve the data processing in order to preserve large-scale structures.

\begin{figure}
\begin{center}
\includegraphics[width=0.49\textwidth]{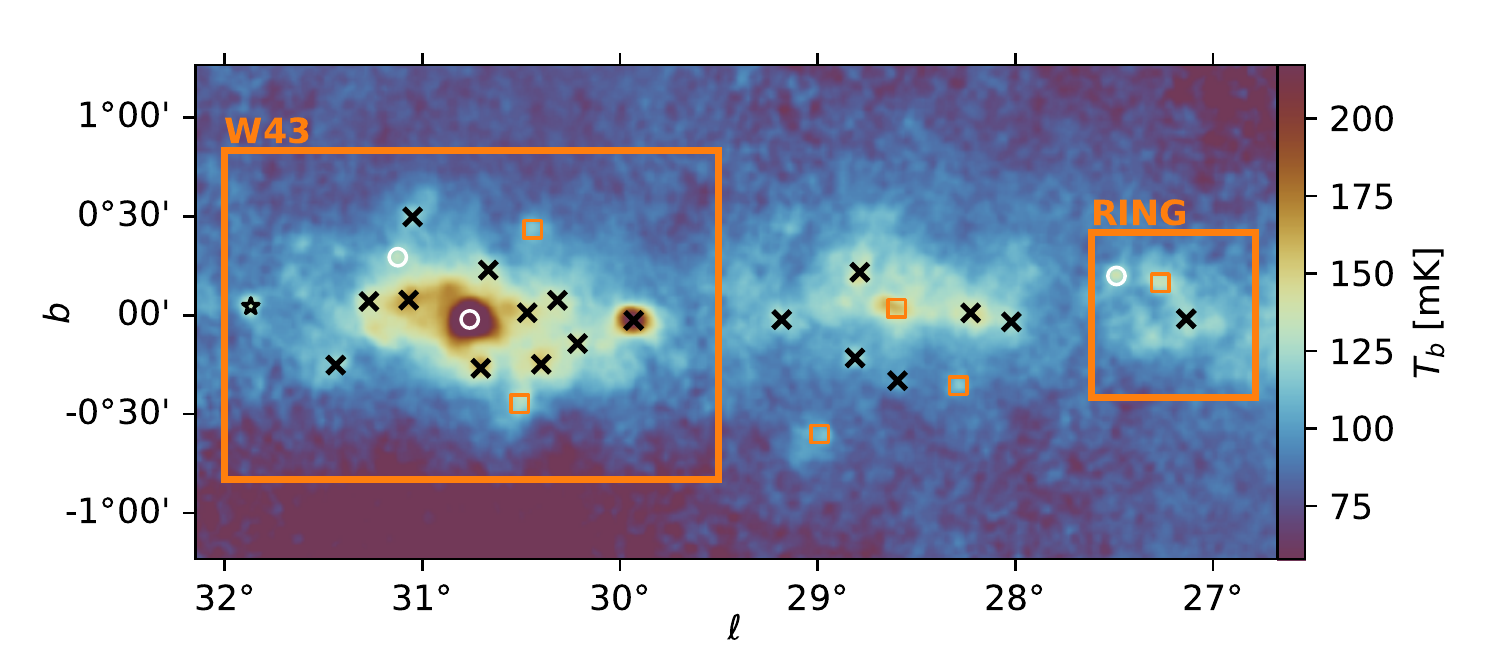}
\caption{COMAP 30\,GHz map of W43 showing the locations of sources extracted: H\textsc{ii} regions found to fit best to a non-AME model (\textit{white circles}), H\textsc{ii} regions found to best fit a model with AME (\textit{orange squares}), SNRs (\textit{black stars}), and other sources detected but discounted from an SED analysis (\textit{black crosses}). We also label the W43 region and ring structure (referred to in \S\ref{sec:G027}) as \textit{orange boxes}.}
\label{fig:W43_close}
\end{center}
\end{figure}

In the following sections we will discuss several initial analyses of the COMAP Galactic plane survey. In \S\ref{sec:results_correlation} we will compare the pixel brightnesses of the diffuse emission at 30\,GHz with surveys at 5 and 353\,GHz. In \S\ref{sec:results_uchii} we discuss the general contribution of UCH\textsc{ii} regions to the total emission observed at COMAP frequencies using data from the CORNISH survey UCH\textsc{ii} catalogue. In \S\ref{sec:results_hii} we discuss nine H\textsc{ii} regions selected from the COMAP Galactic plane survey and look for evidence of AME.  In \S\ref{sec:results_snr} we look at the fitted SEDs of six known supernovae (SNR) over the full $20^\circ < \ell < 40^\circ$ range shown in Fig.\,\ref{fig:COMAP-hits+map}. Finally, in \S\ref{sec:results_rrls} we present preliminary detections of radio recombination lines from both compact and diffuse ionized gas within the COMAP band within the area marked W43 in Fig.~\ref{fig:W43_close}.

\subsection{Correlation with Other Surveys}\label{sec:results_correlation}

We begin the analysis with an initial look at the correlation of the diffuse emission seen at 30\,GHz shown in Fig.~\ref{fig:COMAP-hits+map} with the 5\,GHz Parkes map \citep[primarily tracing free-free emission;][]{Calabretta2014} and the 353\,GHz \textit{Planck} map \citep[tracing dust emission;][]{Planck2018I}. We will derive the average spectral index and Pearson correlation coefficient of the diffuse emission between each frequency pair. If AME is present in the diffuse emission at 30\,GHz we will expect to see a rising spectrum between 5--30\,GHz that is greater than that predicted for free-free emission ($\alpha_{5\--30} > -0.1$), and significant correlation between the 30\,GHz and 353\,GHz data. We account for the possible filtering on large-scales discussed earlier by including a conservative 15\,\% calibration uncertainty on the COMAP data (due to the loss of flux density on large scales discussed in \S\ref{sec:results_comap}).

Before comparing the pixel brightnesses we first smooth all the datasets to a common resolution of $5^{\prime}$ and change the pixel size to match the resolution of the maps to reduce pixel-to-pixel correlations in the noise. We then mask all pixels with a pixel brightness of less than 1\,MJy\,sr$^{-1}$ at 5\,GHz; the mask area is shown by the lowest contour  in Fig.~\ref{fig:COMAP-hits+map}. We also mask the very brightest sources that exceed 6\,MJy\,sr$^{-1}$ at 30\,GHz. The mask ensures we have good S/N in the majority of pixels and mitigates against low-level large-scale modes which are not well constrained away from the Galactic plane.

\begin{figure}[htbp]
\begin{center}
\includegraphics[width=0.8\columnwidth]{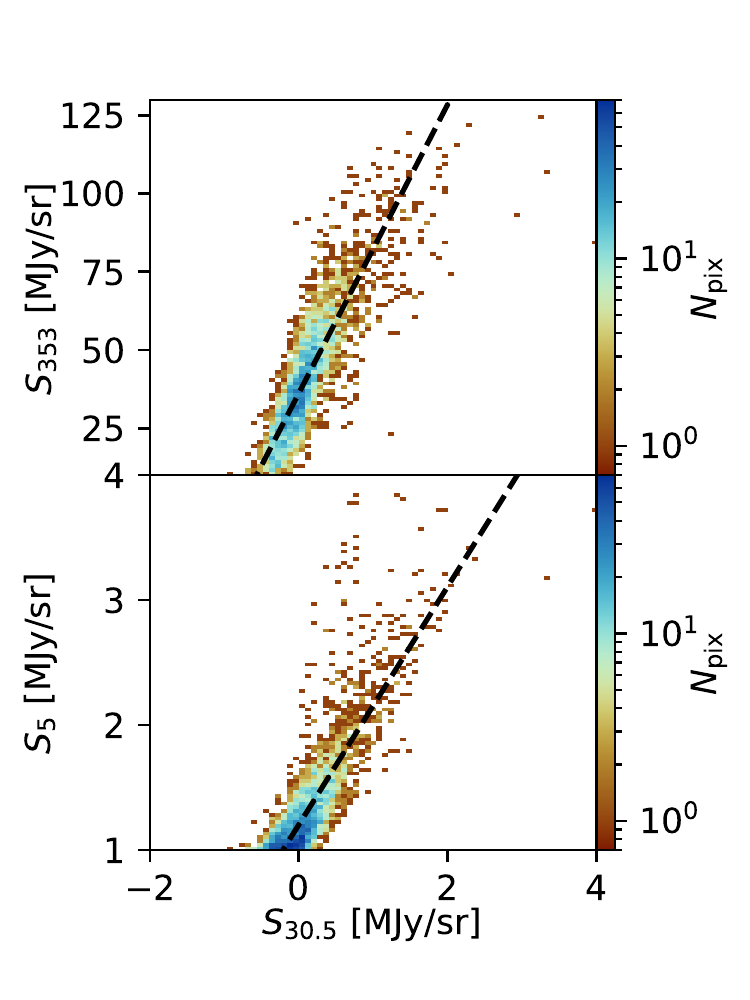}
\caption{\change{Binned T-T plots of p}ixel brightnesses of the COMAP $30$\,GHz data with the Parkes $5$\,GHz (\textit{top}) and \textit{Planck} $353$\,GHz (\textit{bottom}) data. The \textit{dashed} line shows the best-fitting linear model\change{, the parameters of which are discussed in \S\ref{sec:results_correlation}}.}
\label{fig:COMAP-corr}
\end{center}
\end{figure}

In Fig.~\ref{fig:COMAP-corr} we compare the pixel brightness in the 5\,GHz and 353\,GHz maps with COMAP 30\,GHz data. We can see there is a strong correlation between the 30\,GHz data and both 5\,GHz and 353\,GHz emission. The Pearson correlation coefficient between each frequency pair is consistent: $r_{5\--30}=0.72$, $r_{30\--353}=0.68$, and $r_{5\--353}=0.70$. Strong correlations between all three bands are not  unexpected near the Galactic plane as the complex dust and gas dynamics of these regions leads to significant mixing of ISM phases. 

The dashed lines in Fig.~\ref{fig:COMAP-corr} show the best-fit linear relationships between the pixel brightnesses. We convert the fitted gradients into the spectral index between pairs of frequencies by 

\begin{equation}
  \alpha = \frac{\mathrm{log}\left(\mathrm{d}S_{\nu_0}/\mathrm{d}S_{\nu_1}\right)}{\mathrm{log}\left(\nu_0/\nu_1\right)},
\end{equation}
where $\alpha$ is the spectral index between frequencies $\nu_0$ and $\nu_1$. 

The spectral index that we measure between 5\,GHz and 30\,GHz is $\alpha_{5\--30} = 0.035 \pm 0.017$, which is inconsistent with the expected free-free spectrum spectral index of $\alpha_\mathrm{ff} \approx -0.1$ \citep{Draine2011} at the level of $7.9\sigma$. It is possible for a rising spectrum between 5--30\,GHz to be due to optically thick free-free emission from embedded UCH\textsc{ii} regions, but as discussed in \S\ref{sec:results_uchii}, we find that on average UCH\textsc{ii} regions have a negligible impact on these angular scales at 30\,GHz. We therefore interpret this excess as due to AME, probably in the form of spinning dust. We can estimate what the fraction of the emission measured at 30\,GHz is due to AME by

\begin{equation}
    \eta_\mathrm{AME} = 1 - \frac{(\nu_1/\nu_0)^{-0.1}}{(\nu_1/\nu_0)^{\alpha_{5\--30}}},
\end{equation}
where $\nu_1 = 30.5$\,GHz, and $\nu_0=5$\,GHz. We find a fractional AME of $\eta_\mathrm{AME} = (22 \pm 2)$\,\%. There have been several previous estimates of the AME fraction around 30\,GHz. \citet{PlanckEarlyXXI} used a template fitting method to estimate the AME fraction at 28.4\,GHz to be ($25\pm5$)\,\% across all longitudes, but found particular longitudes to be as high as 100\,\%, however such high AME fractions are likely due to modeling errors in complex regions. \citet{PlanckIntXXIII} also performed a template fitting analysis but made use of RRL data over the range $5^\circ < \ell < 60^\circ$ and found an average AME fraction of ($45\pm1$)\,\%. While \citet{PlanckIntXV} found the average AME fraction of individual bright AME regions to be ($50\pm2$)\,\%. 

In future work we will expand this analysis to be a full correlation analysis where we jointly fit for emission components at 30\,GHz using templates in a similar way to what has been done at lower resolutions  \citep[e.g.,][]{Davies2006}. However, since the free-free emission and thermal dust emission are so strongly correlated ($r_{5\--353}=0.70$) such an analysis will require careful consideration of these correlations. 

\subsection{UCH\textsc{ii} Regions}
\label{sec:results_uchii}

As described in \S\ref{sec:fit_strategy}, the SED fitting assumes that free-free emission may be fit by a single optically thin and spatially smooth component. Therefore any contribution from younger, compact H\textsc{ii} regions with rising spectra and higher turnover frequencies may erroneously be classified as AME. These young compact H\textsc{ii} regions can be further classified as ultra- (UCH\textsc{ii}) and hyper-compact (HCH\textsc{ii}) H\textsc{ii} regions depending upon their electron density and emission measure \citep[e.g.,][]{Churchwell2002,Kurtz2002}. The most compact objects (UCH\textsc{ii}/HCH\textsc{ii}) can turnover at frequencies of tens of GHz \citep{Kurtz2002}, although the flux density of individual objects will typically be small ($\sim$mJy). In this section we describe the average contribution of UCH\textsc{ii} regions to the emission at 30\,GHz.

To account for these contributions, one would ideally like to survey at high frequency (e.g., 30\,GHz) \textit{and} high angular resolution ($\sim$arcsec) to allow each source to be accounted for (each source will typically be quite weak - typically 10s-100s\,mJy level). Unfortunately whilst COMAP does have the required frequency coverage, it does not have the angular resolution necessary to perform such a search, however we may consult external catalogues to search for compact sources that may influence our findings. The most appropriate survey to-date is the 5\,GHz CORNISH VLA radio survey \citep{Hoare2012} of the northern Galactic plane.  We used the UCH\textsc{ii} catalog from the $5$\,GHz CORNISH survey \citep{Purcell2012, Kalcheva2018}. The CORNISH survey catalogs $\sim$3000 sources in total at $5$\,GHz with $1\farcs5$ resolution, \change{54 of which lie within our survey area between $20^{\circ}<\ell<40\,^\circ$. Considering that we report $30$\,GHz excess fluxes of $\approx1$\,Jy, we would typically require $\sim 100$ UCH\textsc{ii} regions to account for the excess flux observed.}

The catalog is estimated to  be  $90\,\%$  complete  to  point sources  at $3.9\,$mJy but is less  complete  for  extended  emission, resolving out sources larger than $14^{\prime\prime}$. However, these extended H\textsc{ii} regions, will, by definition, have lower densities and therefore lower turnover frequencies, i.e., they will not be UCH\textsc{ii} regions and are therefore less likely to contribute to the excess emission at $\sim30$\,GHz.

All the sources in the CORNISH UCH\textsc{ii} catalog were classified based on their morphological and positional similarity with IR counterparts in the $8\,\mu$m band of the GLIMPSE survey \citep{Kalcheva2018}. Proximity to IR clusters and dust lanes was used to distinguish UCH\textsc{ii} regions from other objects meeting the above criteria such as massive young stellar objects and planetary nebulae. 

We have assessed the contribution from the \change{54} cataloged UCH\textsc{ii} sources to the fitted AME and found it to be negligible. However, we must also consider the potential for sources that were not included in the catalog to make a significant contribution. These sources fall into two main categories: i) any compact H\textsc{ii} sources that appeared in the parent CORNISH catalog but were not detected in GLIMPSE (referred to as ``IR quiet``) and so could not be classified as UCH\textsc{ii} regions, and ii) sources falling below the completeness level of the parent CORNISH catalog.
    
Of the CORNISH IR quiet sources 80$\%$ have flux densities of $\leq13$\,mJy. Therefore, we conservatively estimate the contribution of sources \change{missing from the catalogue} by assuming a true completeness level of the UCH\textsc{ii} catalog of $13$\,mJy (over 3 times the point source completeness limit of the CORNISH survey).

\cite{Yang2021} observed 116 young H\textsc{ii} regions that have rising spectra from $1$ to $5$\,GHz \change{(5 in our survey area)}, and found 20 sources with turnover frequencies $>5$\,GHz \change{(4 in our survey area)}, with maximum and mean turnover frequencies of $16.7$\,GHz and $9.7$\,GHz respectively. If we consider a source with a $5$\,GHz flux density of $13$\,mJy and assume the source to have the maximum turnover frequency ($\nu_t=16.7$\,GHz) reported by \cite{Yang2021}, its $30$\,GHz flux density is just $\sim13$\,\% of the mean excess ($\Delta S_{30}$ in Table~\ref{tab:HII}). This contribution decreases significantly to $\sim4$\,\% if we assume the source to have the mean turnover frequency ($\nu_t=9.7$\,GHz). 

Given the rarity of UCH\textsc{ii} regions with such high turnover frequencies, we conclude that sources missing from the CORNISH UCH\textsc{ii} catalog do not make a significant contribution to the reported AME in this paper. The largest contribution in any of the sources initially found by our analysis is seen to be $\approx 4\%$ although most sit well below this contribution. Nevertheless, as we expand the COMAP survey, there might be some lines-of-sight where UCH\textsc{ii} regions, particularly if there is a cluster within a molecular cloud, could be a major fraction of the total flux density. In Section~\ref{sec:results_hii} we also provide details on the UCH\textsc{ii} region contributions to each of the six AME sources. 

\subsection{H\textsc{ii} region SEDs}
\label{sec:results_hii}

At frequencies of $\nu < 100$\,GHz H\textsc{ii} regions are dominated by free-free emission, which has a well defined spectral shape \citep{Draine2011} that at frequencies between $5 < \nu < 100$\,GHz is often optically thin with a spectral index of $\alpha_\mathrm{ff} \approx -0.1$. There are several examples of AME being detected in H\textsc{ii} regions \citep{Watson2005,Dickinson2007,PlanckIntXV}, as well as observations associating the AME with the swept up circumstellar material at the boundaries of H\textsc{ii} regions \citep{Tibbs2010,Tibbs2012,Cepeda-ArroitaEtAl2021}. However, not all H\textsc{ii} regions show evidence of AME \citep{Scaife2007,Scaife2008}. In this section we will discuss nine sources selected from the COMAP Galactic plane survey that are coincident with H\textsc{ii} sources, six of which we find to have a $> 3\sigma$ 30\,GHz excess that is indicative of AME, and three which show no evidence for AME.

\begin{figure*}[htbp]
\begin{center}
\includegraphics[width=0.33\textwidth]{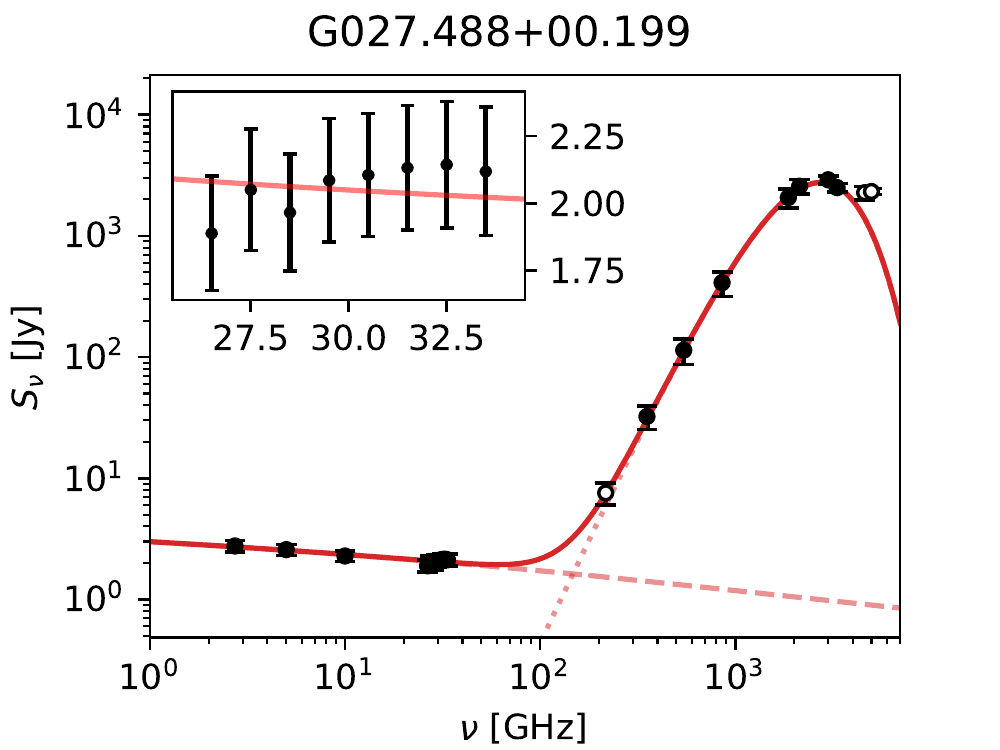}
\includegraphics[width=0.33\textwidth]{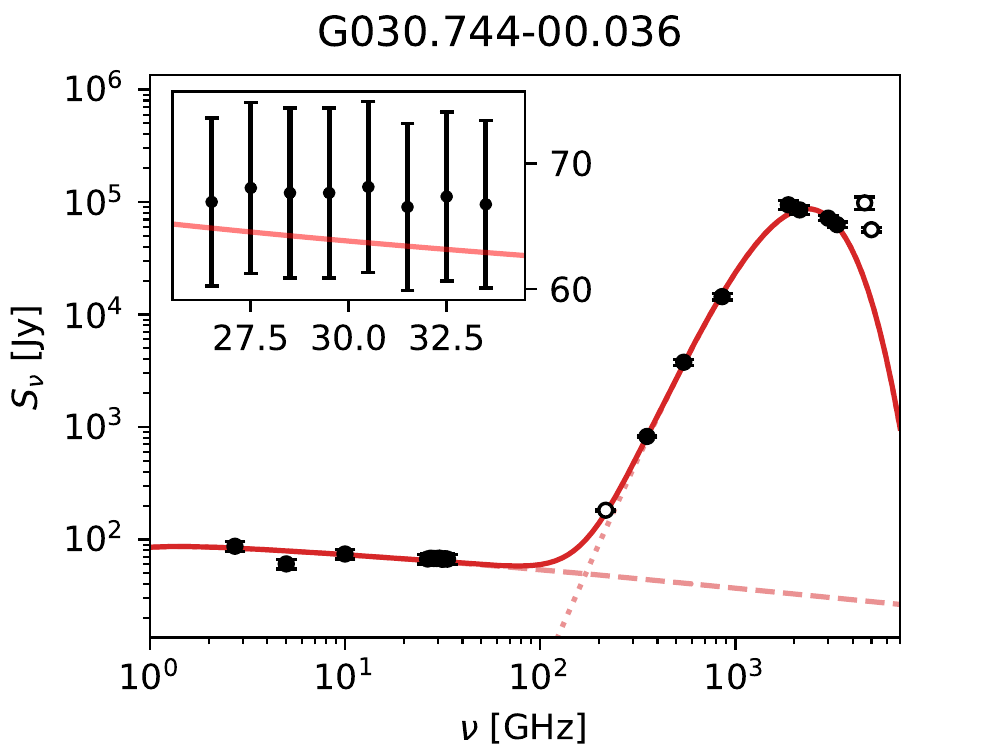}
\includegraphics[width=0.33\textwidth]{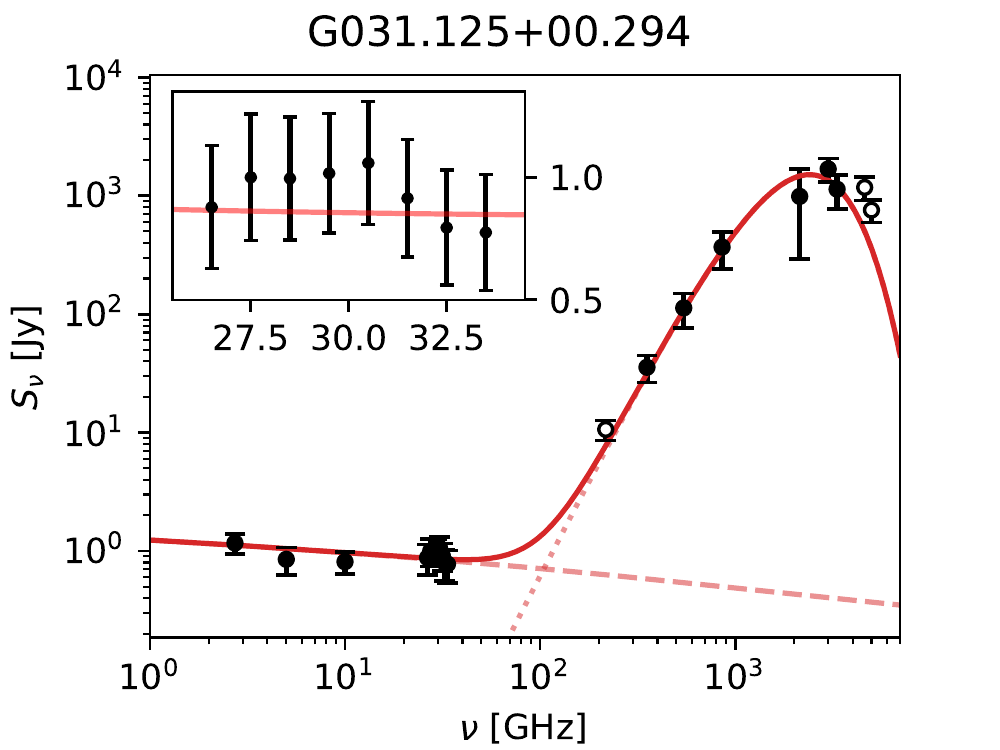}
\caption{SEDs of three H\textsc{ii} regions without an AME \change{detection.} Filled circles are used in the fit while open circles were not. The free-free and thermal dust components are shown as dashed and dotted lines respectively. The total emission model is shown as the thick red line. The cutout shown in the upper left of all plots shows the COMAP in-band 26--34\,GHz spectrum.}
\label{fig:sed_no}
\end{center}
\end{figure*}

\begin{figure*}[htbp]
\begin{center}
\includegraphics[width=0.33\textwidth]{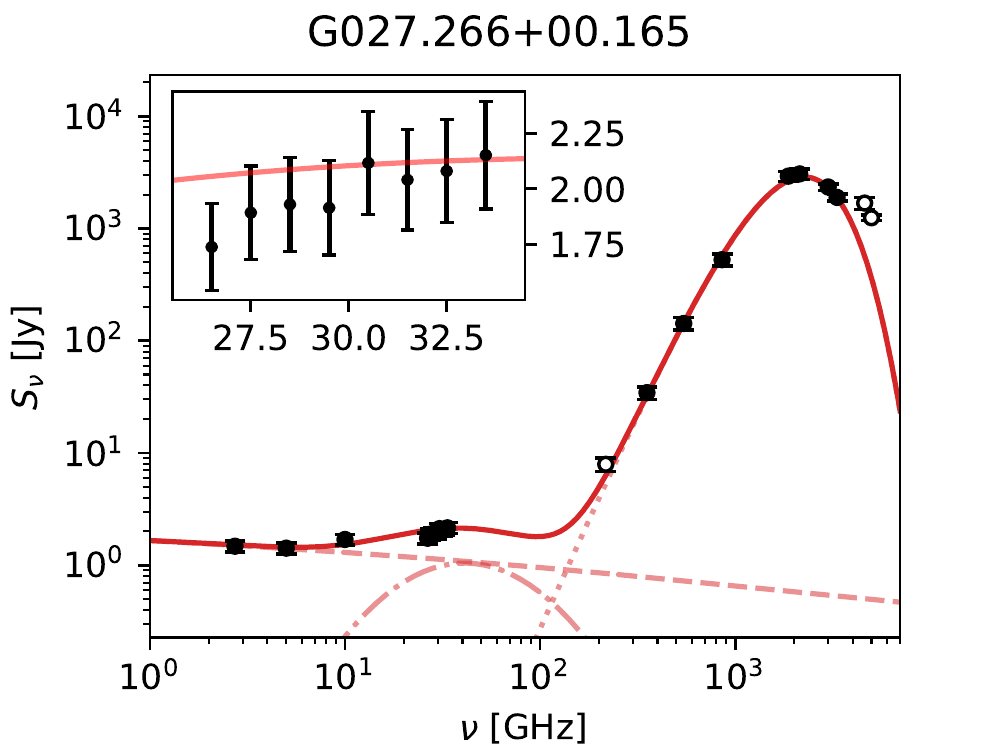}
\includegraphics[width=0.33\textwidth]{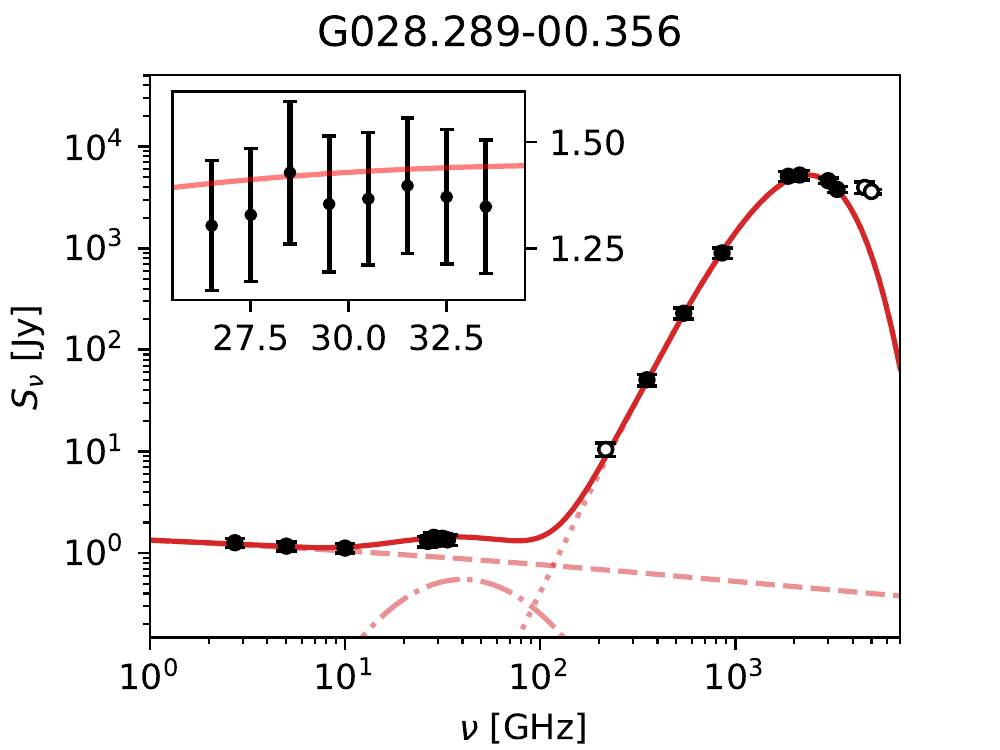}
\includegraphics[width=0.33\textwidth]{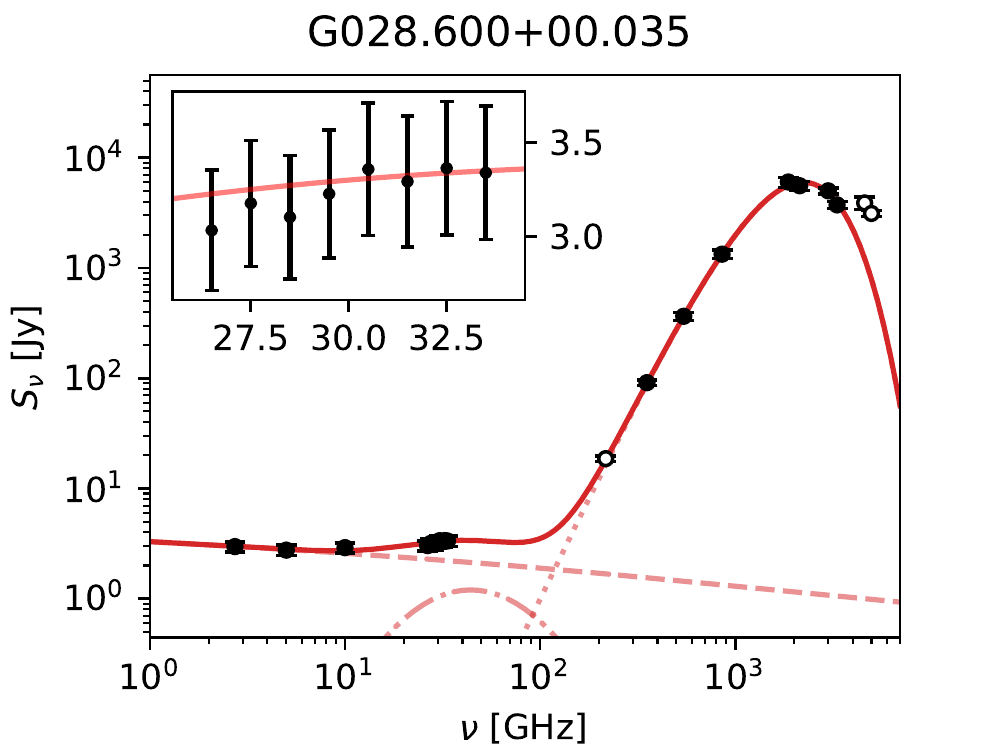}
\includegraphics[width=0.33\textwidth]{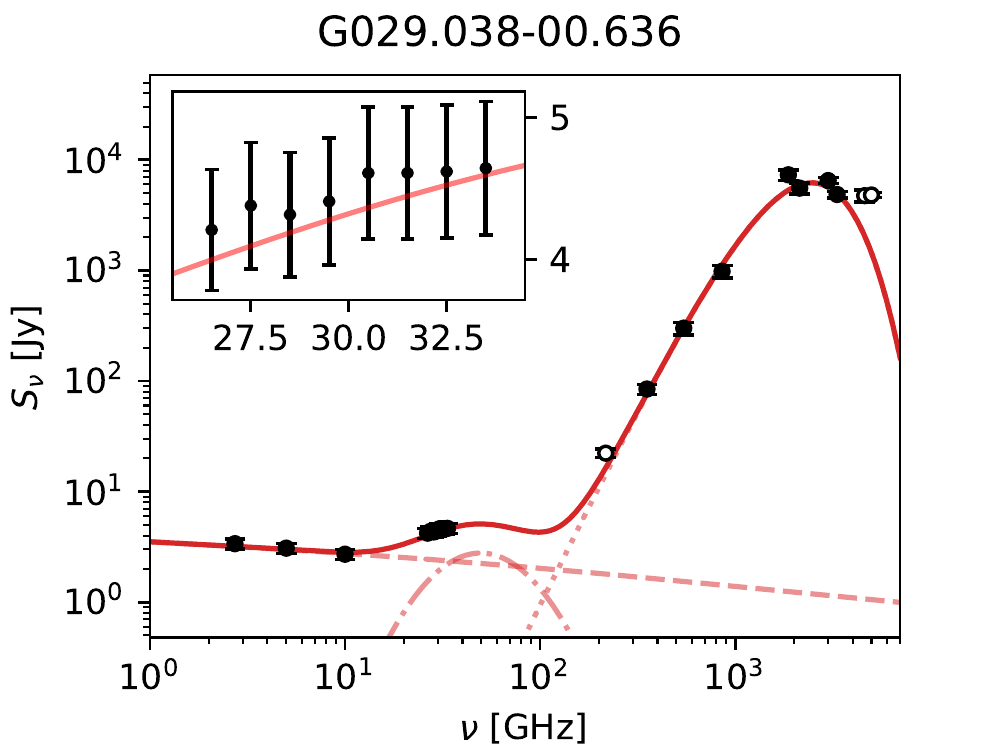}
\includegraphics[width=0.33\textwidth]{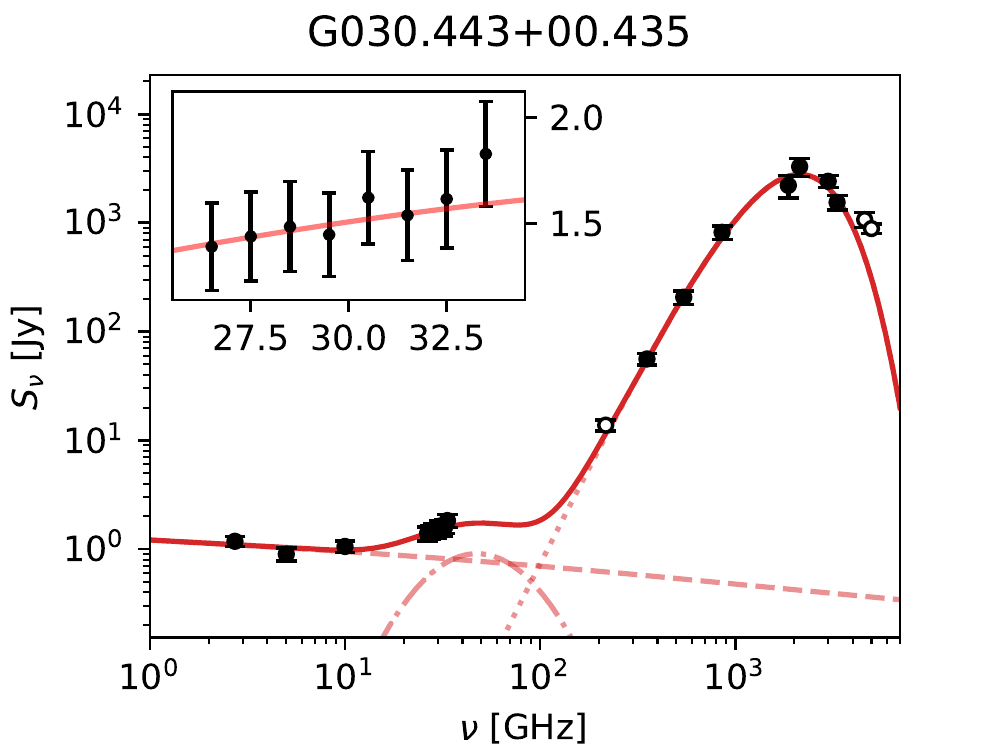}
\includegraphics[width=0.33\textwidth]{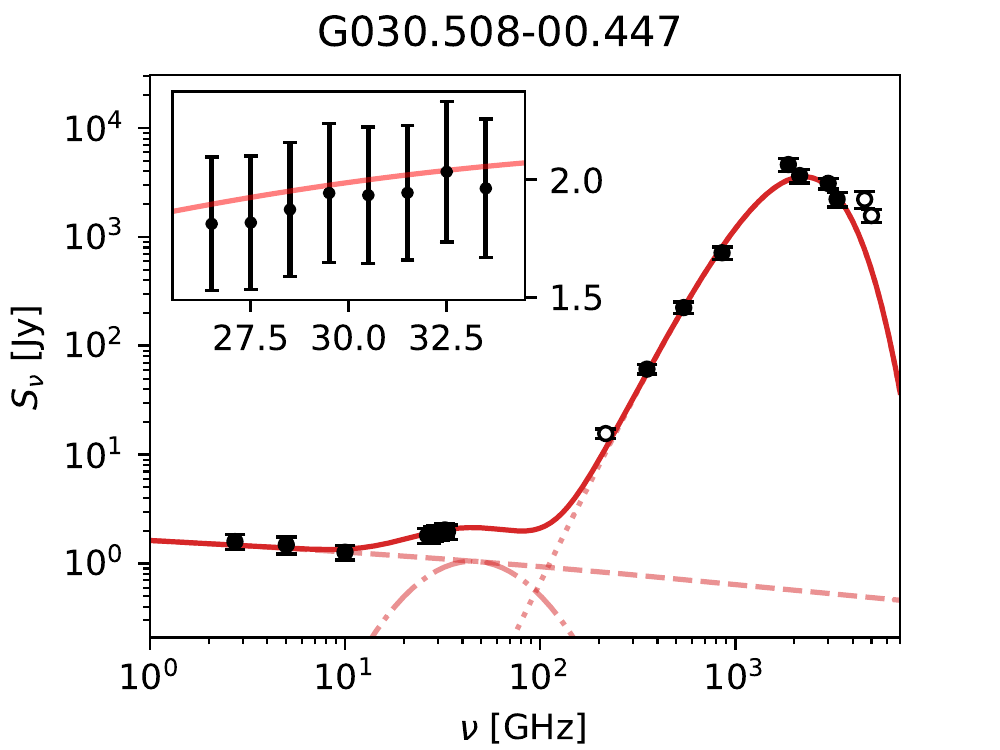}
\caption{SEDs of six H\textsc{ii} regions that show a significant AME detection. Filled circles are used in the fit while open circles were not. The free-free, spinning dust, and thermal dust components are shown as dashed, dash-dot and dotted lines respectively. The total emission model is shown as the thick red line. The cutout shown in the upper left of all plots shows the COMAP in-band 26--34\,GHz spectrum.}
\label{fig:sed_yes} 
\end{center}
\end{figure*}

In Fig.\,\ref{fig:sed_no} we show the best-fit spectral energy distribution (SED) models to the non-AME sources that are dominated by optically thin free-free emission at 30\,GHz, and Fig.~\ref{fig:sed_yes} shows the model fits to the six sources that exhibited a significant excess at 30\,GHz above the fitted free-free emission component. %The robustness of the model fits is tested by looking at the reduced $\chi^2/\mathrm{d.o.f.}$ statistic. 

In \change{Table~\ref{tab:uchiichi2}} we give the  $\chi^2/\mathrm{d.o.f.}$ of each region, as well as the relative contribution to the $\chi^2/\mathrm{d.o.f.}$ for data between 2.7--10\,GHz ($\chi^2_\mathrm{lo}$), 26--35\,GHz ($\chi^2_\mathrm{COMAP}$), and 100--4000\,GHz ($\chi^2_\mathrm{hi}$). For most regions we find \mbox{$0.5 < \chi^2/\mathrm{d.o.f.} < 1.5$}. Estimates where the $\chi^2/\mathrm{d.o.f.}$ is less than unity is due to the background correlation uncertainties not being accounted for and the correlation between the calibration of each survey. The assumption that the calibration uncertainties we give in Table~\ref{tab:surveys} are uncorrelated, results in an over estimate in the uncertainty for some regions. The worst fitting regions are G030.744$-$00.036 (the core of W43) and G029.038$-$00.636 (RCW\,175) with $\chi^2/\mathrm{d.o.f.} \sim 3$. For G030.744$-$00.036 the large $\chi^2/\mathrm{d.o.f.}$ is driven by the low frequency data, specifically the flux density measured at  5\,GHz using the Parkes survey is underestimated relative to the two other low frequency surveys. We expect that this is due to issues with calibration of the Parkes data in this region, likely due to significant sidelobe pickup from the nearby, bright diffuse emission that surrounds G030.744$-$00.036 \change{(although further investigation would be required to prove this)}. The other region with a high $\chi^2/\mathrm{d.o.f.}$ is G029.038$-$00.636, in this case we find that the issue \change{is} associated with just the \textit{Akari} data. A comparison between the IRAS and \textit{Akari} data did not reveal any clear morphological differences (e.g., no point sources had been subtracted from the reprocessed IRAS data in the region), suggesting a systematic calibration issue with \textit{Akari} data along this line of sight. 

We estimate the emission measure ($\mathrm{EM} \equiv \int n_e^2 \mathrm{d}l$) of each source from the fitted amplitude of the free-free model described in \S\ref{sec:sed_fitting}. We find that the aperture-averaged $\mathrm{EM}$ of the sources in our sample are in the range $5000$--$20000$\,pc\,cm$^{-6}$, which is typical for classical H\textsc{ii} regions \citep{Kurtz2002}. The exception is W43 itself which has a higher $\mathrm{EM}$ of 400,000\,pc\,cm$^{-6}$, comparable to previous estimates of the brightest (and densest) central source with $\mathrm{EM}=720,000$\,pc\,cm$^{-6}$ \citep{Downes1980}, and as expected given that it is a massive star formation region. As all of the sources are either approximately the size of the COMAP beam or slightly extended we are not significantly underestimating the true $\mathrm{EM}$ of the sources due to beam dilution, however the value does represent an average over the measured aperture area.

\change{We also fit models for thermal dust emission which give dust temperatures ($T_d$ in Equation 12) in the range $23$--$30$\,K. These measurements agree with \citet{Anderson2012b} who report dust temperatures of $25.3\pm2.4$\,K averaged across whole HII regions. Notably these dust temperatures are higher than those reported by \citet[19.4$\pm$1.3\,K;][]{PlanckIntXLVIII}, who mainly consider the high-latitude sky, where the dust is known to be colder.}

As well as fitting full SED models we also fit for the spectral index of the emission measured within the COMAP frequency band (26--34\,GHz). In-band spectral indices are given in Table~\ref{tab:HII} ($\alpha_{26\--34}$). We find that there may be evidence that the AME regions are more likely to have a rising spectra within the COMAP band, this will become clearer in the future as we measure more AME and non-AME regions in the COMAP Galactic plane survey.

The peak frequencies of the candidate AME sources are given in Table~\ref{tab:HII}. We find that our peak frequencies are higher than the  25--30\,GHz range found previously \citep[e.g.][]{PlanckIntXV,Planck2015XXV}, with a mean value of $\nu_\mathrm{AME} \sim\!40$\,GHz. We tested several different priors and found that changes in the peak frequency for different priors were all consistent within the uncertainties. The apparent higher than expected AME peak frequency must therefore be driven by the lack of data between 40 and 100\,GHz, meaning we cannot constrain the downturn in the AME spectrum. Although the peak frequencies we measure are generally high, they are still consistent within 1$\sigma$ with the expected AME peak frequency of $\sim\!30$\,GHz.

The AME widths that we measure are between $0.55 < w_\mathrm{AME} < 0.85$, similar to the range of widths measured in the $\lambda$-Orionis ring \citep{Cepeda-ArroitaEtAl2021}. In general, we find that the AME width is not well constrained for these sources for the same reason we cannot easily constrain the peak frequency (i.e., no data between 40 and 100\,GHz). However, we do notice that there tends to be a relationship between the width of the AME spectrum and the measured in-band spectral index, with steeper in-band spectra leading to narrower widths implying we do have some constraining power from the COMAP data alone. However, at present this is not a clear relationship, for example G027.27.266$+$00.165 has a steep in-band spectrum but also a wide $w_\mathrm{AME}$, which is mostly driven by an excess of emission seen in the 10\,GHz Nobeyama data. 

In order to obtain an AME emissivity (brightness per unit column density of dust) for each source ($\epsilon_\nu$) we use the excesses and $\tau_{353}$ values reported in Table~\ref{tab:HII}. We convert $\tau_{353}$ to a column density via $(8.4\pm3.0)\times10^{-27}$ cm$^2$/H as reported by \citet{Planck2013XI} for the whole sky. The uncertainties are dominated by this conversion factor. These values make up the last column in Table~\ref{tab:HII} and show a clear difference between non-AME sources (most of which are consistent with zero emissivity) and the AME regions, which typically have emissivities of the order $10^{-17}$ at 30\,GHz. 

In Fig.~\ref{fig:sed_no} we show the SEDs of the H\textsc{ii} regions in which we found no evidence of AME. The SEDs are fitted with a two-component free-free plus thermal dust model. The region G027.488+00.199 is  located near to the H\textsc{ii} region G027.266+0.165 and just outside of the ring structure marked in Fig.~\ref{fig:W43_close}. The region H\textsc{ii} G031.125+00.294 lies near the exterior of the larger W43 complex. Both G027.488+00.199 and G031.125+00.294 are classical H\textsc{ii} regions; they contain a number of infrared dark clouds \citep{Peretto2009} but are not associated with any known stellar clusters or ionizing main sequence stars \citep{Reed2003,Bica2019}. The region G030.744$-$00.036 is centered on W43, one of the Galaxy's most active star formation regions \citep[e.g.,][]{Nguyen2011}; using  a central $8^{\prime}$ aperture we measure  a flux density of $\sim100$\,Jy, approximately 20\,\% of the flux density of the entire region \citep{Irfan2015,Genova_SantosEtAl2017}.

\begin{figure*}
    \centering
    \includegraphics[width=0.98\textwidth]{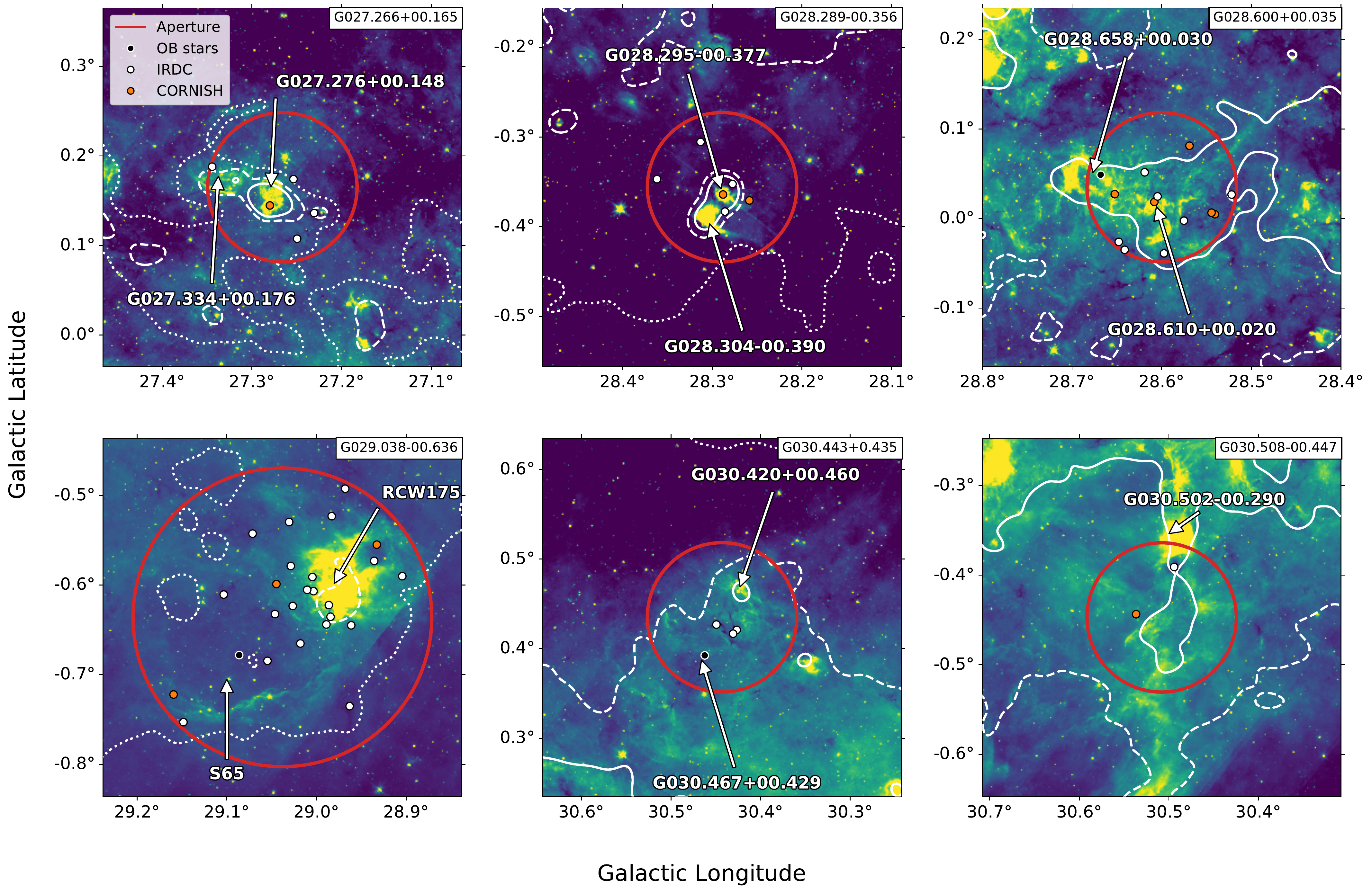}
    \caption{\textit{Spitzer} 8\,$\mu$m IRAC images of each H\textsc{ii} region where we find significant AME. The contours are from the 250\,$\mu$m Hi-GAL data at 1000 (\textit{dotted} line), 2000 (\textit{dashed} line), and 3000\,MJy\,sr$^{-1}$ (\textit{solid} line). The \textit{red filled circle} marks the aperture used to measure the SED. The \textit{white} markers show the locations of IRDCs from \citet{Peretto2009}. Sources from the CORNISH survey $>\!7\sigma$ catalogue are shown as \textit{orange} markers \citep{Purcell2012}. \textit{Black} markers indicate O/B stars \citep{Reed2003}.}
    \label{fig:ame_src1}
\end{figure*}

In the following we will discuss the properties of each extracted H\textsc{ii} region in more detail. In Fig.~\ref{fig:ame_src1} we show \textit{Spitzer} 8\,$\mu$m GLIMPSE \citep[image;][]{Churchwell2009}  and \textit{Herschel} 250\,$\mu$m Hi-Gal survey \citep[contours;][]{Molinari2016} data of each region in which we detect AME. All of these regions are sites of active or recent star formation. The annotations are the H\textsc{ii} region designations defined in \citet{AndersonEtAl2014}; we also mark the locations of bright 5\,GHz sources from the CORNISH survey \citep{Purcell2012}, the locations of infrared dark clouds \citep[IRDC;][]{Peretto2009}, and any known O/B stars \citep{Reed2003}.

\subsubsection{G027.266$+$00.165} \label{sec:G027}

The G027.266$+$00.165 region contains two H\textsc{ii} regions  at a distance of $\sim\!2.6$\,kpc \citep{AndersonEtAl2014}, and is coincident with a larger ring of diffuse emission (marked in Fig.~\ref{fig:W43_close}) that is visible at $\nu < 40$\,GHz and in the far-infrared IRAS data. We can see in Fig.~\ref{fig:ame_src1} that the H\textsc{ii} region \change{G027.276+00.148} is centered in the aperture. The second H\textsc{ii} region G027.334+00.176 appears to exhibit a photodissociation region (PDR) along its lower boundary \change{(identified by the arc of emission subtending $\approx0.1\,^\circ$ seen at $8$\,$\mu$m on the edge of the IRDC)}, which may be associated with the nearby embedded cluster \citep{Bica2019}; PDRs have previously been found to be sources of AME \citep{Casassus2008}. 

We find that the AME emissivity of this region is the highest out of all the AME detections ($10.1\pm3.8$\,Jy/sr\,cm$^{2}$/H; Table~\ref{tab:HII}), which could potentially be linked to the presence of the PDR. In 250\,$\mu$m data we see that the cold dust emission \change{is} largely concentrated around the main central H\textsc{ii} region and the nearby IRDCs. The very large dust columns associated with IRDCs could be associated with regions of high AME emissivity \citep[e.g.,][]{Ysard2011}, which is another possible explanation for the regions high emissivity. Finally, it is interesting that we do not detect any AME in the nearby H\textsc{ii} region \change{G027.488$+$00.199}, despite being physically close together and \change{sharing} similar environmental conditions. However, there is one key difference between the two regions, G027.266$+$00.165 is coincident with the diffuse background emission of the ring marked in Fig.~\ref{fig:W43_close} while  \change{G027.488$+$00.199 is separated from it},  which suggests the AME may be associated with the diffuse background emission and not necessarily the H\textsc{ii} region itself. 

The 10\,GHz data point in Fig.~\ref{fig:sed_yes} is seen to be rising above the free-free emission spectrum. One \change{possible} explanation for this rise is simply a systematic issue with 10\,GHz data in this region. Another \change{possibility is that it is} evidence for an optically thick UCH\textsc{ii} region with a turn over frequency of $\sim\!10$\,GHz. In \S\ref{sec:results_uchii} we discussed the typical UCH\textsc{ii} contributions at 30\,GHz but there are a number that are much brighter than average, one of which happens to be within the G027.266$+$00.165 region. To determine whether this specific UCH\textsc{ii} is optically thick at 30\,GHz we use the deconvolved source size ($5.5^{\prime \prime}$) and integrated flux density ($S_5 = 0.428$\,Jy) provided in the CORNISH catalogue \citep{Purcell2012}, and calculate the brightness temperature at 5\,GHz. We find that the brightness temperature of this UCH\textsc{ii} region is $< 1000$\,K at 5\,GHz, which is much less than the typical electron temperature of H\textsc{ii} regions ($\sim10^4$\,K), which implies that the region is optically thin at 5\,GHz, and therefore cannot explain the excess we see at 30\,GHz.

\subsubsection{G028.289$-$00.356}

 This region contains two H\textsc{ii} regions G028.295$-$00.377 and G028.304$-$00.390 at distances of 3.3\,kpc and 4.8\,kpc respectively \citep{AndersonEtAl2014}. Both H\textsc{ii} regions contain embedded clusters \citep{Bica2019}. The source is not resolved by the COMAP beam. As well as the two H\textsc{ii} regions, the aperture contains three IRDCs, and two candidate UCH\textsc{ii} sources identified by the CORNISH survey \citep{Yang2021}. This region does not have any clear evidence of large PDRs, or any known O/B stars. We estimate the two UCH\textsc{ii} regions to be both optically thin at 5\,GHz, and therefore not contributing to the observed AME excess seen by COMAP.
 
 % This region seems to have more in common with the non-AME regions, and is also the lowest excess AME, but the AME fractions are more or less the same.
 
 %One of the CORNISH UCH\textsc{ii} regions has a 5\,GHz flux density of $S_5 = 0.55$\,Jy and is extended with a geometric mean diameter of $4.4^{\prime\prime}$, which gives a mean brightness temperature of $T_b^{\rm ff} \sim 1600$\,K implying the region is optically thin and cannot explain the excess at 30\,GHz. The second UCH\textsc{ii} region is not resolved in the CORNISH survey but has a flux density of $S_5=18.34$\,mJy, which, even if it was optically thick, is not bright enough to contribute significantly to the observed AME excess at 30\,GHz. 
 
\subsubsection{G028.600+00.035}

The region G028.600+00.035 lies close to the Galactic plane, and is embedded within a much larger complex that is clearly visible at 30\,GHz in Fig.~\ref{fig:W43_close}. It contains the H\textsc{ii} region G028.610+00.020 near the center of the aperture, and another H\textsc{ii} region G028.658+00.030 that is found near to the edge, both H\textsc{ii} regions are at a distance of 6.2\,kpc \citep{AndersonEtAl2014}. G028.658+00.030 contains the high-mass star LS\,IV\,-03\,8 \citep{Reed2003}. This region is potentially the most active star formation region of the six AME candidate sources, containing five UCH\textsc{ii} candidates from the CORNISH survey, and six IRDC \citep{Peretto2009}. The COMAP emission shows that this source is slightly extended ($6.9^\prime$; Table~\ref{tab:HII}), and the peak in the 30\,GHz emission is not coincident with either of the H\textsc{ii} regions. There is a bright filamentary structure in the 8\,$\mu$m data, that might be evidence of an extended PDR. This region is also the brightest AME region in the 250\,$\mu$m data, implying it has the largest cold dust column density of all six AME regions. 

Of the six UCH\textsc{ii} regions five are extended all with estimated mean surface brightness temperatures of $T_b^{\rm ff} < 1000$\,K implying they are optically thin and therefore are fitted by the optically thin free-free emission model. The one unresolved UCH\textsc{ii} region in the CORNISH catalogue has a flux density of $3.23$\,mJy and contributes a negligible flux density at 30\,GHz even if it is optically thick.

\subsubsection{G029.038$-$00.636}

This region contains two H\textsc{ii} regions: RCW175 \citep{Rodgers1960}, and S65 \citep{Sharpless1959}---both regions have a kinematic distance of $\sim3$\,kpc \citep{Lee2012}. The region has been studied several times before and is a well known source of AME \citep{Dickinson2009,Tibbs2012,Battistelli2015}. RCW175 is an extended source $\approx 10^{\prime}$ in size, and is clearly visible in the 8\,$\mu$m data. The region contains an embedded cluster \citep{Bica2019}, a high density of IRDCs \citep[14 in total;][]{Peretto2009}, and a large number of protostellar objects \citep{Kuhn2021} making this object one of the more active star forming regions in our sample. 

The other H\textsc{ii} region S65 \citep{Sharpless1959} is associated with the O/B star ALS\,19303 \citep{Reed2003} that is not visible at 8\,$\mu$m but is at radio frequencies \citep{Paladini2003} indicating it is likely a Str\"{o}mgren sphere \change{\citep[spherical region of ionized gas surrounding a young O/B type star;][]{Stromgren1939}}. We can see that there is a bubble of swept up circumstellar material around the boundary of S65, and that there is evidence of a PDR visible at 8\,$\mu$m along the lower edge of the region.

The region is resolved in the COMAP data, with a geometric mean diameter of $12.7^\prime$. Interestingly we find that the peak in the COMAP 30\,GHz map is not coincident with either H$\textsc{ii}$ region but instead peaks between the two---though it is not clearly coincident with any particular feature such as the PDR seen near S65.

There are two candidate UCH\textsc{ii} regions within the CORNISH survey for RCW175. One of which is has a flux density of $S_5=4.65$\,mJy, resulting in a negligible contribution at 30\,GHz. While the other has a flux density of $S_5=20.21$\,mJy, which if we use the maximum turn-over frequency of 16.7\,GHz given in \S\ref{sec:results_uchii} gives an upper limit of $S_{30}=0.2$\,Jy at 30\,GHz---much less than the observed 30\,GHz excess ($\Delta S_{30} = 1.97\pm0.14$).

\subsubsection{G030.443+0.435}

This region is situated above the larger diffuse W43 complex as seen in Fig.~\ref{fig:W43_close} and contains a single bright H\textsc{ii} region: S66/RCW176 \citep{Sharpless1959,Rodgers1960}, that is referenced as G030.420$+$00.460 in the \textit{WISE} H\textsc{ii} catalogue at a distance of 3.6\,kpc \citep{AndersonEtAl2014} and potentially contains an embedded cluster \citep{Bica2019}. The \textit{WISE} catalogue cites a second H\textsc{ii} region G030.467$+$00.429 which is not associated with any emission at 8\,$\mu$m but is visible in the radio \citep{Paladini2003}; the region may be ionized by the coincident high-mass star LS IV -02 16 \citep{Reed2003}. Unlike other AME regions discussed, there are no associated  compact 5\,GHz sources in the CORNISH catalogue implying there is no current high-mass star formation occurring. Several IRDC are found to be clustered around the center of the aperture, and are coincident with the 30\,GHz COMAP emission. These are likely the source of the AME and would be ideal candidates for high frequency follow-up observations.

\subsubsection{G030.508$-$00.447}

G030.508$-$00.447 is situated below the W43 complex as seen in Fig.~\ref{fig:W43_close}. The region is dominated by a diffuse spur of emission that is correlated at all frequencies implying that the gas and dust within the ISM are highly mixed in this region. This is the only region in this sample that does not contain an H\textsc{ii} region within the COMAP aperture. G030.502$-$00.290 lies just outside the aperture at a distance of 7.3\,kpc, but it may not be directly associated with the dust spur. There is a single IRDC cloud with a high peak 8\,$\mu$m opacity of $\tau_{8\mu\mathrm{m}} = 5.68$ \citep{Peretto2009} near to the edge of the aperture, but it appears that most of the emission at 8\,$\mu$m has at least some absorption, indicative of a high density cloud. Although there is no compact H\textsc{ii} source, the region is still undergoing active star formation with 15 candidate young stellar objects (most of which are early stage protostellar objects: Class I/II) mostly concentrated around the IRDC although several are scattered throughout the spur \citep{Kuhn2021}.

We find that there is only one faint CORNISH source with a 5\,GHz flux density of $S_5 = 6$\,mJy, which we find to be optically thin and makes a negligible contribution to the observed COMAP 30\,GHz emission.

%\thomas{
%\begin{itemize}
%\item AME found in $?$\,\% of sources
%\item Tau-353 to excess ratio distribution
%\item $nu_{AME}$ distribution
%\item excess vs. latitude analysis
%\item UCHII analysis to validate 
%\end{itemize}}
\small
\setlength{\tabcolsep}{3pt}
\begin{table*}
\caption{Fitted SED parameters for H\textsc{ii} sources analyzed with their associated geometric mean widths from source extraction ($\theta$), showing in band spectral indices ($\alpha_{26-34}$), fitted parameters are given for models discussed in \S\ref{sec:sed_fitting}, model excess emission ($\Delta S_{30}$) and emissivity at $30$\,GHz ($\epsilon_\nu$) calculated from the excess and optical depth at 353\,GHz. For non-AME sources we present 2$\sigma$ upper limits on the last two parameters: $30$\,GHz excess and emissivity.}
\label{tab:HII}
\begin{tabular*}{\textwidth}{l c c ccc c ccc c c}
\hline\hline
\multicolumn{2}{c}{Source} & In-Band & \multicolumn{3}{c}{Thermal Dust} & Free-Free & \multicolumn{3}{c}{AME Lognormal} \\
\hline
\multicolumn{1}{c}{Name} & $\theta$ & $\alpha_\textrm{26-34}$ & $T_d$ & $\beta$ & $\log_{10}(\tau_{353})$ & $\mathrm{EM} \times 10^{-3}$ & $A_\textrm{AME}$ & $\nu_\textrm{AME}$ & $w_\textrm{AME}$ & $\Delta S_{30}$ & $\epsilon_\nu\times10^{18}$\\
& [$^\prime$] & & [K] & & & [pc cm$^{-6}$] & [Jy] & [GHz] &  & [Jy] & [Jy/sr cm$^2$/H] \\
\hline
G027.488$+$00.199 & \change{$4.5$} & $0.5^{\pm0.1}$ & $30^{\pm 3}$ & $1.50^{\pm0.24}$ & $-3.79^{\pm0.07}$ & $12.7^{\pm0.4}$ &  ...  &  ...  &  ...  & $<0.13$ & $<1.25$ \\
G030.744$-$00.036 & \change{$6.7$} & $-0.0^{\pm0.1}$ & $22^{\pm 1}$ & $2.10^{\pm0.08}$ & $-2.19^{\pm0.02}$ & $398^{\pm14}$ & ...  &  ...  &  ...  & $<1.49$ & $<0.5$ \\
G031.125$+$00.294 & \change{$3.9$} & $-0.8^{\pm0.5}$ & $27^{\pm 4}$ & $1.30^{\pm0.29}$ & $-3.71^{\pm0.13}$ & $5.3^{\pm0.4}$ &  ...  &  ...  &  ...  & $<0.21$ & $<2.8$ \\
\hline
G027.266$+$00.165 & \change{$5.9$} & $0.8^{\pm0.1}$ & $22^{\pm 1}$ & $2.02^{\pm0.16}$ & $-3.56^{\pm0.05}$ & $7.1^{\pm0.7}$ & $1.0^{\pm0.5}$ & $41^{\pm17}$ & $0.81^{\pm0.21}$ & $0.98^{\pm0.10}$ & $10.1^{\pm3.8}$ \\
G028.289$-$00.356 & \change{$3.8$} & $0.1^{\pm0.1}$ & $23^{\pm 1}$ & $2.03^{\pm0.16}$ & $-3.40^{\pm0.05}$ & $5.7^{\pm0.6}$ & $0.6^{\pm0.4}$ & $40^{\pm16}$ & $0.73^{\pm0.23}$ & $0.52^{\pm0.07}$ & $9.2^{\pm3.5}$ \\
G028.600$+$00.035 & \change{$6.6$} & $0.4^{\pm0.1}$ & $23^{\pm 1}$ & $1.79^{\pm0.12}$ & $-3.14^{\pm0.03}$ & $14.0^{\pm1.0}$ & $1.2^{\pm0.3}$ & $44^{\pm13}$ & $0.71^{\pm0.20}$ & $1.06^{\pm0.11}$ & $3.3^{\pm1.2}$ \\
G029.038$-$00.636 & \change{$12.1$} & $0.4^{\pm0.1}$ & $27^{\pm 2}$ & $1.70^{\pm0.19}$ & $-3.35^{\pm0.05}$ & $17.5^{\pm1.2}$ & $3.6^{\pm1.0}$ & $47^{\pm10}$ & $0.56^{\pm0.13}$ & $1.97^{\pm0.14}$ & $3.0^{\pm1.1}$\\
G030.443$+$00.435 & \change{$6.1$} & $1.0^{\pm0.2}$ & $22^{\pm 2}$ & $1.66^{\pm0.24}$ & $-3.34^{\pm0.06}$ & $5.2^{\pm0.5}$ & $0.9^{\pm0.3}$ & $47^{\pm13}$ & $0.59^{\pm0.20}$ & $0.70^{\pm0.08}$ & $3.2^{\pm1.2}$\\
G030.508$-$00.447 & \change{$4.9$} & $0.4^{\pm0.1}$ & $23^{\pm2}$ & $1.73^{\pm0.23}$ & $-3.35^{\pm0.05}$ & $6.9^{\pm0.8}$ & $1.1^{\pm0.4}$ & $45^{\pm14}$ & $0.66^{\pm0.22}$ & $0.89^{\pm0.10}$ & $8.3^{\pm3.1}$\\
\hline
\end{tabular*}

\end{table*}

\setlength{\tabcolsep}{6pt}
\normalsize

\setlength{\tabcolsep}{3pt}
\begin{table}
\caption{Summary of reduced $\chi^2/\mathrm{d.o.f.}$ values determined for the 9 sources in Table\,\ref{tab:HII}. $\chi^2/\mathrm{d.o.f.}$ values are given for the whole fitted dataset alongside the  contributions from data points with $\nu<26$\,GHz ($\chi^2_\mathrm{lo}$), for the 8 COMAP bands ($\chi^2_\mathrm{COMAP}$) and for datasets with $\nu>34$\,GHz ($\chi^2_\mathrm{hi}$).}
\label{tab:uchiichi2}
\centering
\begin{tabular}{lcccc}
\hline\hline
\small
Name & $\chi^2/\mathrm{d.o.f.}$ & $\chi^2_\mathrm{lo}$ & $\chi^2_\mathrm{COMAP}$ & $\chi^2_\mathrm{hi}$ \\
\hline
G027.266$+$00.165 & $1.53$ & $0.18$ & $0.97$ & $0.39$ \\
G030.744$-$00.036 & $3.42$ & $1.9$ & $0.46$ & $1.06$ \\
G031.125$-$00.294 & $1.01$ & $0.34$ & $0.38$ & $0.29$ \\
\hline
G027.488$+$00.199 & $0.53$ & $0.02$ & $0.34$ & $0.17$ \\
G028.289$-$00.356 & $0.55$ & $0.04$ & $0.34$ & $0.17$ \\
G028.600$+$00.035 & $0.61$ & $0.08$ & $0.15$ & $0.38$ \\
G029.038$-$00.636 & $2.99$ & $0.59$ & $0.08$ & $2.33$ \\
G030.443$+$00.435 & $1.31$ & $0.38$ & $0.25$ & $0.68$ \\
G030.508$-$00.447 & $1.52$ & $0.11$ & $0.11$ & $1.3$ \\
\hline
\end{tabular}
\end{table}
\normalsize
\setlength{\tabcolsep}{6pt}

\setlength{\tabcolsep}{1pt}
\begin{table}
\caption{Summary of measurements of 6 SNRs detected in the current COMAP survey.}
\label{tab:snr}
\centering
\begin{tabular}{lcccc}
\hline\hline
\small
Name                &$\theta$       &$S_{30}$         &$\alpha_{2.7-30}$  &$\alpha_{26-34}$   \\
                    &[$^{\prime}$]  &[Jy]               &                   &       \\
\hline
G021.5$-$0.9        &5              &$5.52^{\pm0.56}$   &$-0.07^{\pm0.06}$  &$+0.14^{\pm0.09}$ \\
G021.8$-$0.6\,(Kes69)&20             &$5.99^{\pm0.69}$  &$-0.77^{\pm0.06}$  &$-1.30^{\pm0.30}$ \\ 
G31.9$+$0.0\,(3C\,391)   &7$\times$5     &$3.08^{\pm0.34}$   &$-0.60^{\pm0.06}$ &$-1.05^{\pm0.22}$\\
G34.7$-$0.4\,(W44)   &35$\times$27   &$41.0^{\pm 4.2}$              &$-0.54^{\pm0.0.06}$             &$-0.50^{\pm0.11}$  \\
G35.6$-$0.4         &15$\times$11   &$6.2^{\pm1.1}$     &$-0.34^{\pm0.08}$                   &$-2.5^{\pm0.5}$           \\
G039.2$-$0.3\,(3C\,396) &8$\times$6     &$6.83^{\pm0.73}$   &$-0.25^{\pm0.06}$  &$+0.11^{\pm0.11}$\\
\hline
\end{tabular}
\end{table}
\normalsize
\setlength{\tabcolsep}{6pt}

\subsection{Supernova Remnants}
\label{sec:results_snr}

Few supernova remnants (SNRs) are readily detectable at higher radio frequencies ($\nu \gtrsim 30$\,GHz), owing to their typically steeply falling with frequency synchrotron spectrum ($\alpha \!\sim\! -0.7$) and increased background. However, the brightest SNRs and those with a flatter spectrum ($\alpha > -0.5$) are detectable. Those with flat radio spectra tend to have a filled-centre or mixed morphology (rather than just a shell) and are often associated with a pulsar wind nebula (PWN); the Crab nebula is one of the best examples \citep[for a review, see][]{Dubner2015}.  

A wide range of spectra have been observed, particularly over a \change{wide frequency} range, where a different population of electron energies are probed. The differences are thought to be due to various intrinsic and extrinsic sources of energy injection/losses, resulting in a variety of spectra and morphologies \citep{Dejan2014,Dubner2015}. Thus, high frequency observations are useful for understanding the energy losses/re-acceleration as well as characterising non-synchrotron components from free-free and AME. Indeed, a number of SNRs have been reclassified as H\textsc{ii} \citep[e.g.,][]{Dokara2021}.

We have inspected the entire 30\,GHz COMAP survey to-date at the location of the 33 known SNRs from the most recent version\footnote{http://www.mrao.cam.ac.uk/surveys/snrs/} of the Green supernova catalogue \citep{Green2019} over the longitude range $l=20^{\circ}$--$40^{\circ}$. Most of these SNRs show no obvious source in the COMAP map, or, are confused with other nearby sources and background emission. There are 6 SNR sources of interest, three of which are very extended ($\gtrsim 10^{\prime}$) and three which are compact or only slightly larger than the beamwidth ($\gtrsim 5^{\prime}$). We also looked at the positions of new SNR candidates from \cite{Dokara2021} but no obvious detection was possible.

\begin{figure*}
    \centering
    \gridline{
    \fig{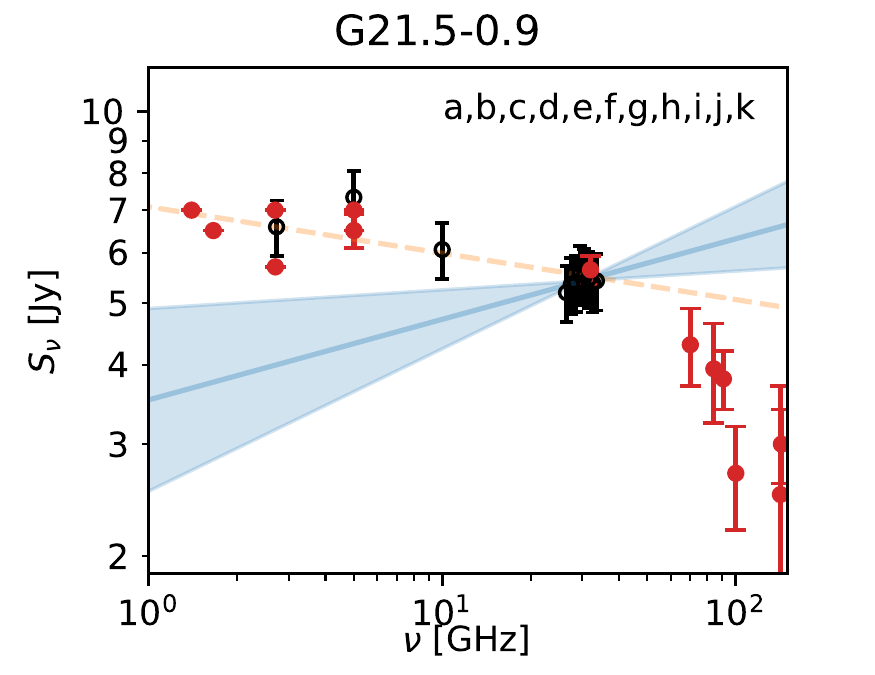}{0.33\textwidth}{}
    \fig{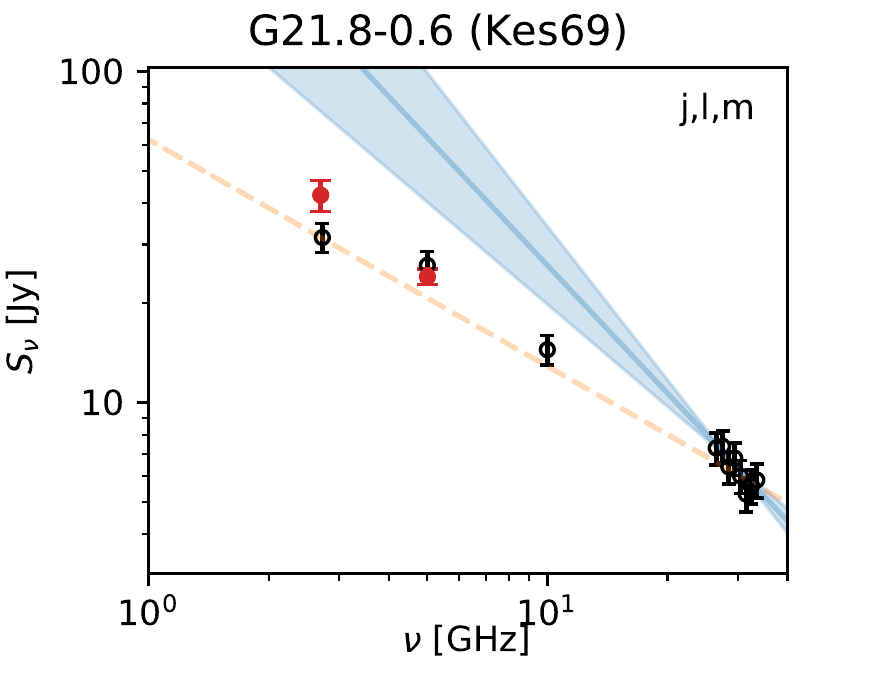}{0.33\textwidth}{}
    \fig{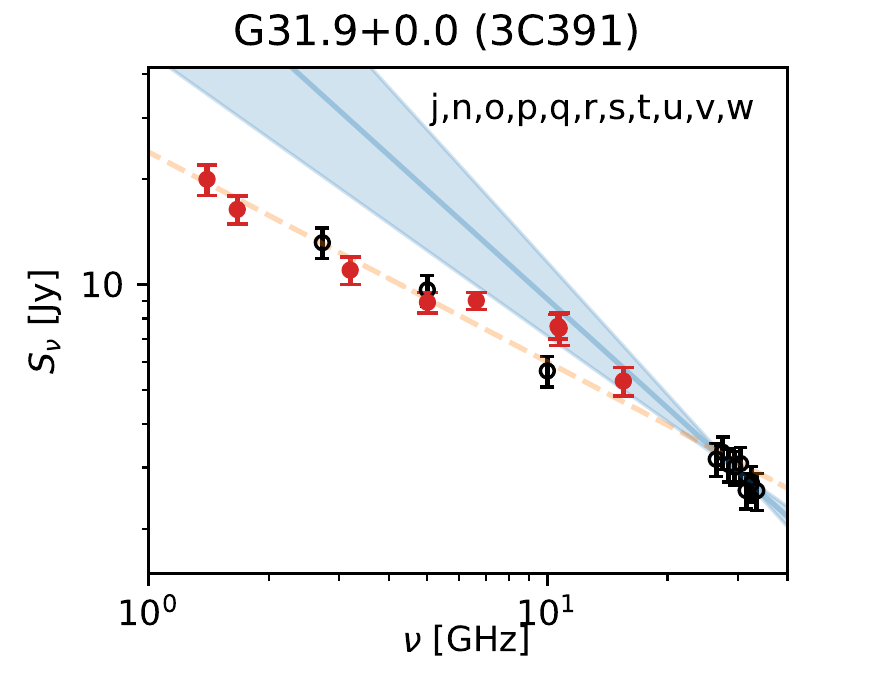}{0.33\textwidth}{}
    }
    \gridline{
    \fig{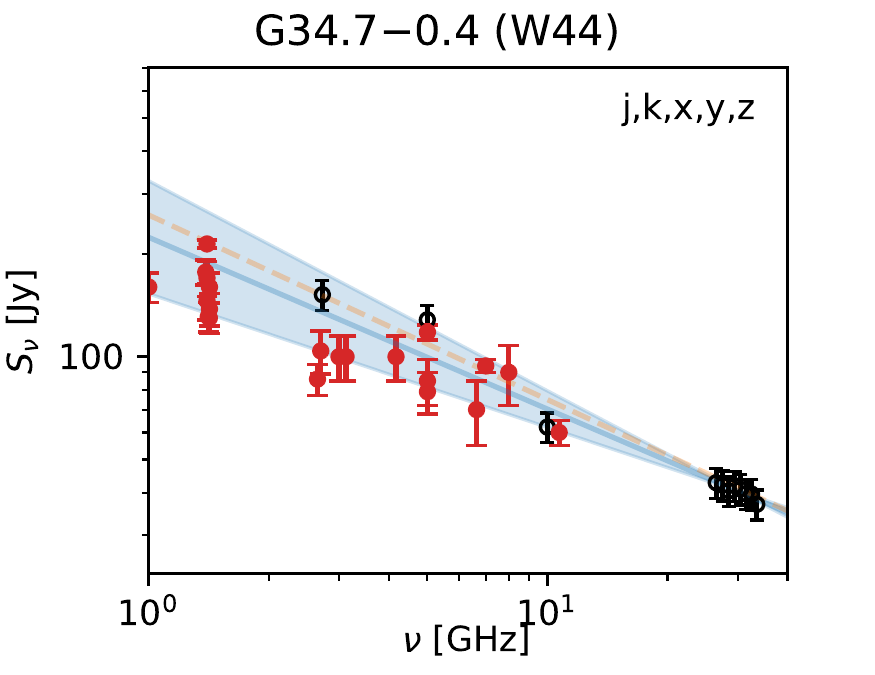}{0.33\textwidth}{}
    \fig{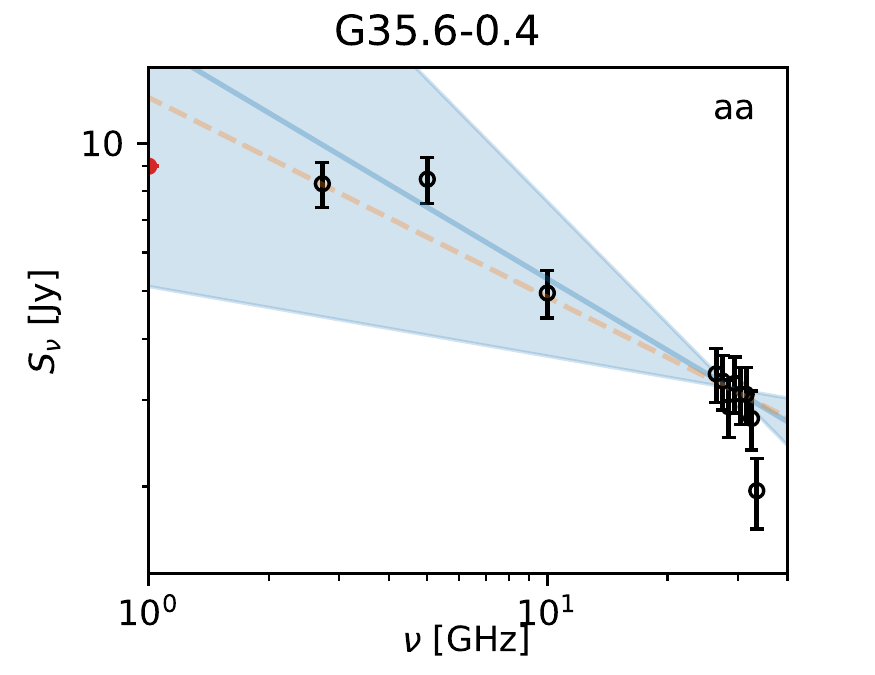}{0.33\textwidth}{}
    \fig{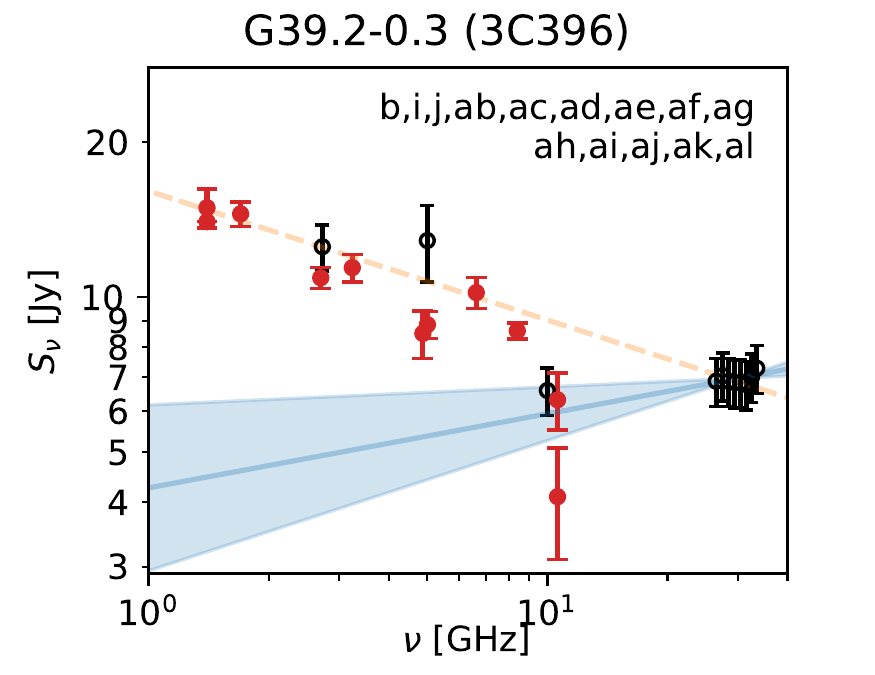}{0.33\textwidth}{}
    }
    \caption{SEDs of the 6 SNR discussed in \S\ref{sec:results_snr} showing extracted flux densities in \textit{black}, with literature values in \textit{red} \change{(references for these points are given in top-right of each plot)}. Each SED also shows the fitted $2.7$--$30$\,GHz power-law model (\textit{orange dashed} line) and the in-band $26$--$34$\,GHz fit (\textit{blue} line). The in-band spectrum extrapolated to lower/higher frequencies as a power-law, with the \textit{blue} shaded region showing the associated 1$\sigma$ uncertainty. Note that for G35.6$-$0.4 we show the fitted 26--32\,GHz spectral index (i.e., without the last two COMAP bands) due to apparent evidence of a spectral break in the in-band spectrum.\\ \change{\textbf{References:} (a) \citet{Clark1974}; (b) \citet{Altenhoff1970}; (c) \citet{Milne1969}; (d) \citet{Goss1970}; (e) \citet{Reifenstein1970}; (f) \citet{Morsi1987a}; (g) \citet{Salter1989}; (h) \citet{Salter1989b}; (i) \citet{Kassim1992}; (j) \citet{Sun2011}; (k) \citet{PlanckIntXXXI}; (l) \citet{Velusamy1974}; (m) \citet{Green1997}; (n) \citet{Dulk1972}; (o) \citet{Caswell1971}; (p) \citet{Artyukh1969}; (q) \citet{Condon1971}; (r) \citet{Kesteven1968}; (s) \citet{Pauliny-Toth1966}; (t) \citet{Bridle1971}; (u) \citet{Chaisson1974}; (v) \citet{Goss1979}; (w) \citet{Moffett1994}; (x) \citet{Clark1974}; (y) \citet{Castelletti2007}; (z) \citet{Egron2017}; (aa) \citet{Green2009}; (ab) \citet{Dickell1975}; (ac) \citet{Becker1976}; (ad) \citet{Becker1987}; (ae) \citet{Dulk1975}; (af) \citet{Shavers1970}; (ag) \citet{kellerman1969}; (ah) \citet{ReichEtAl1984}; (ai) \citet{Downes1981}; (aj) \citet{Day1970}; (ak) \citet{Hughes1969}; (al) \citet{Cruciani2016}}}
    \label{fig:snr_sed}
\end{figure*}

We briefly discuss them in turn and include our flux density measurement at 30\,GHz along with the spectral index from 2.7\,GHz to 30\,GHz and in-band spectral index fitted separately over the 26--34\,GHz COMAP frequency range. The results are given in Table\,\ref{tab:snr}. For each SNR, we adjust the aperture size to be suitable for the SNR and use a background annulus that is 1.3 to 1.6 times the radius. The SEDs of these six SNRs are presented in Fig.\,\ref{fig:snr_sed}, showing our measurements from the maps available to us as well as values from the literature. \change{Note that in order to stay as consistent with the literature as possible, we use the names (and coordinate conventions) first allocated to each of the 6 SNRs when referring to them.}

\subsubsection{G34.7$-$0.4 (W44)}

We begin with W44 (G34.7$-$0.4, 3C\,392), which is a well-studied SNR residing in and interacting with a giant molecular cloud \citep[e.g.,][]{Rho1994} at a distance of $\approx 3$\,kpc. W44 is clearly visible in the COMAP 30\,GHz map (Fig.\,\ref{fig:COMAP-hits+map}) and looks similar in appearance to low frequency radio maps \citep[e.g.,][]{Castelletti2007}. It has a distorted morphology over approx $0.\!\!\degree5$ but with a clear shell across to the SE. Previous estimates of the radio spectral index are around $\alpha=-0.4$ to $-0.3$ between 0.02--1\,GHz. However, recent observations by the Sardinia Radio Telescope (SRT) at 1.5 and 7\,GHz \citep{Egron2017}, and 21.4\,GHz \citep{Loru2019} have shown that the spectrum is \textit{steepening} dramatically at high ($\gtrsim 10$\,GHz) frequencies, suggesting that there is spectral break in the cosmic ray energy spectrum that has already been observed with gamma-ray observations of W44 \citep{Ackermann2013}. 

Using the higher frequency data from COMAP we can help confirm the energy of the spectral break, and help constrain the mechanism behind cosmic ray production in W44. Given the complexity of the region and its large angular size (which allows spectral variations across the source to be assessed), we will present a detailed analysis in a future work (Harper et al., in prep.). Nevertheless, our initial analysis finds a flux density of $S_{30}=41.0 \pm 4.2$\,Jy, which is consistent with a power-law model from lower frequencies (Fig.\,\ref{fig:snr_sed}). The best-fitting spectral index is $\alpha_{2.7-30}=-0.54 \pm 0.06$, with in-band spectral index $\alpha_{26-34}=-0.50 \pm 11$. Therefore, the integrated SED appears to be well-approximated by a power-law over the range $1$--30\,GHz. We do not see evidence for the level of steepening observed by \cite{Loru2019}, who measure spectral indices above 7\,GHz of $\alpha \approx -1$. \cite{Loru2019} note that their 21.4\,GHz flux density of $25\pm3$\,Jy falls far short of fits to lower frequency data; our power-law fit predicts $49.1\pm0.6$\,Jy at 21.4\,GHz. \change{However, low frequency SRT data from \cite{Egron2017} are consistent with our power-law relation, with fluxes of 214$\pm$6 and 94$\pm$4\,Jy at 1.5 and 7\,GHz, respectively.} Given the high angular resolution of the SRT data, and large-scale negatives apparent in their map (see their figure 5), it is possible that there could be some flux loss in their data, which could be responsible for an under-estimate of the integrated flux of W44 and steeper indices.

\subsubsection{G021.5$-$0.9}

G021.5$-$0.9 is a relatively compact ($<5^{\prime}$) bright source associated with a PWN \citep{Bietenholz2011}. It is detected with high S/N as a compact source in the COMAP maps. With a $5^{\prime}$ aperture we find $S_{30}=5.52\pm0.56$\,Jy, which is consistent with the 32\,GHz Effelsberg measurements of $5.64\pm0.29$\,Jy \citep{Morsi1987a}. The radio spectrum has been measured to be close to flat $(\alpha=-0.06\pm0.03)$ up to tens of gigahertz, which we confirm ($\alpha_{2.7-30}=-0.07\pm0.06$). The spectrum is known to have a spectral break around 30--40\,GHz \citep{Becker1976,Morsi1987a,Salter1989,Bietenholz2011} with $\alpha=-0.37\pm0.19$ between 32 and 84\,GHz \citep{Salter1989}. Our in-band spectral index is also consistent with a flat spectrum ($\alpha_{26-34}=0.14\pm0.09$) indicating that the spectral break must be above 34\,GHz and must be relatively sharp. This is supported by the data at frequencies $\sim 100$\,GHz (see Fig.\,\ref{fig:snr_sed}).

\subsubsection{G021.8$-$0.6 (Kes\,69)}

G021.8$-$0.6 (Kes\,69) is a diffuse and extended SNR of $20^\prime$ in size \citep[e.g.,][]{Bietenholz2011}. G021.8$-$0.6 has a limb-brightened shell, which is visible in the COMAP survey, while the rest of the shell is 
difficult to discern without background removal. Very few measurements have been made at higher radio frequencies ($>10$\,GHz). 

Fig.\,\ref{fig:snr_sed} shows the SED with data from the literature and our own analysis. At lower frequencies, the spectrum is well-fitted to a power-law up to 10\,GHz with $\alpha=-0.56\pm0.03$ \citep{Sun2011}. Unlike plerionic SNRs \change{(Crab-like, driven by a central pulsar)}, this SNR is dominated by its shell. We do not attempt to measure the total flux density of the source as a whole, but instead focus on the limb-brightened shell using a $10^\prime$ aperture. We find spectral indices of $\alpha_{2.7-30}=-0.77\pm0.06$ and $\alpha_{26-34}=-1.30\pm0.30$. Both these values hint at some spectral ageing above 10\,GHz, increasing with frequency.

\subsubsection{G31.9+0.0 (3C\,391)}

G31.9+0.0 (3C\,391) is a well-known shell-dominated SNR of size  $7^{\prime}\!\times\!5^{\prime}$. It has a well-measured spectrum \citep{Sun2011} that is flat below 1\,GHz. Above 1\,GHz, there is a spectral break \citep{Moffett1994} into a power-law like spectrum with $\alpha=-0.54\pm0.02$ \citep{Sun2011}. Using a $7^{\prime}$ aperture we find $S_{30}=3.08\pm0.34$\,Jy and $\alpha_{2.7-30}=-0.60\pm0.06$. The in-band spectrum is $\alpha_{26-34}=-1.05\pm0.22$. These are both consistent with previous fits, but with an indication of steepening above 30\,GHz. As noted by \cite{Sun2011}, the spectrum appears to be slightly more complicated than a simple power-law; there are hints that the spectrum flattens around 10\,GHz and then steepens again above 30\,GHz in the literature (see Fig.\,\ref{fig:snr_sed}). However, our best-fitting power-law to our own extractions is consistent with a power-law from 2.7\,GHz to 30\,GHz, with only a hint of steepening above 30\,GHz.

\change{A broken power-law trend would be typical of an aging SNR, where high-frequency photons begin to suffer energy loss through radiative cooling. This causes the spectral index to steepen for frequencies greater than that of the break frequency (typically around a few tens of GHz). As this happens, a morphological change occurs, from a shell and towards a composite shape, which may be identified by higher resolution instruments such as the JVLA.}

\subsubsection{G35.6$-$0.4}

G35.6$-$0.4 is an interesting source that was initially classified as an SNR but was then removed following its identification as a thermal source, possibly due to nearby PNe IRAS18554+0203. Using new radio data \cite{Green2009} re-identified the source as an SNR. It is an extended source of $15^{\prime}\!\times\!11^{\prime}$ with a limb-brightened shell. Very few measurements exist for this source (Fig.\,\ref{fig:snr_sed}). \cite{Green2009} measured the radio spectral index of $\alpha=-0.47\pm0.07$. It is detected at high S/N at 2.7\,GHz but is weak at 30\,GHz, barely being detected in the COMAP map. Nevertheless, we use a $15^{\prime}$ aperture to estimate a spectral index and find $\alpha_{2.7-30}=-0.34\pm0.08$, which is consistent with \cite{Green2009}, and supporting the SNR nature of the source. 

At higher frequencies  we find that the in-band spectrum appears to be remarkably steep: $\alpha_{26-34}=-2.5\pm0.5$. However, the bright relative background makes this measurement difficult, hence the large uncertainty. We tried varying the aperture/background annuli sizes and found this trend to be robust. We also notice some additional faint diffuse 30\,GHz emission to the south of the SNR, which looks like interlocking shells, also visible in the 2.7\,GHz data. This has not been observed before and possibly could be related to the SNR. An initial analysis with photometry did not yield robust spectral index to confirm this, but will be investigated in a future work.

\subsubsection{G39.2$-$0.3 (3C\,396)}

Finally, G39.2$-$0.3 (3C\,396) is a relatively well-known compact $8^{\prime}\!\!\times\!\!6^{\prime}$ SNR. A 
review of flux densities is given by \cite{Cruciani2016}. At 33\,GHz they give an updated VSA flux density of $5.20\pm0.33$\,Jy, however, this is after taking into account spatial filtering of the original measurement. Fig.\,\ref{fig:snr_sed} indicates that some measurements, particularly at higher frequencies, are lower than expected compared to a simple power-law model.

With a $10^{\prime}$ aperture we find $S_{30}=6.83\pm0.73$\,Jy, which is more consistent with the original 33\,GHz value from the Very Small Array \citep{Scaife2007} \change{of $(6.64\pm0.3)$\,Jy} . We find a best-fitting power-law from 2.7\,GHz of $\alpha_{2.7-30}=-0.25\pm0.06$, \change{which is slightly flatter} than previous fits at lower frequencies of $\alpha=-0.466\pm0.024$ over 150\,MHz up to tens of GHz \citep{Cruciani2016}. This suggests that the spectrum may be flattening, possibly due to associated free-free or AME components, as originally suggested by \cite{Scaife2007}. Our in-band value ($\alpha_{26-34}=0.11\pm0.11$) is consistent with a flat spectrum, which supports this hypothesis. Such an additional component may be originating in the region affected by the SNR, or it could be a thermal source along the line of sight.

%G$020.0-0.2$. Very few measurements. \cite{Sun2011} gives a brief review 5 GHz flux density of $9.23\pm0.54$\,Jy and $\alpha=-0.06\pm0.04$. No observations above 5 GHz. We can test if this spectrum remains to higher frequencies or indeed if this is not an SNR? \textcolor{blue}{CD: No HDF5 file? There is a detection in the map, although relatively faint. Spectrum looks steeper COMAP in-band ($\alpha \sim -0.8$?). Probably not reliable as is on the edge of the map.}

%G$29.7-0.3$ (Kes\,75). Relatively compact 3\,arcmin. Check central coords! (e.g. 2.7 GHz). Confirm 32 GHz Effelsberg flux density of $1.02\pm0.07$\,Jy \citep{Morsi1987}. Well-defined spectrum (fig.\,3 of \citet{Morsi1987}. Again, wee seem to be low by about 20\%. In-band spectrum also looks flat. Probably too faint to believe without more investigation.

%G$36.6+2.6$. Very few measurements! Mostly Reich 1988/1990. 17x13 arcmin. Very clear at 2.7 GHz and not published? Can we see it at 30 GHz? (maps are saturated) Off the edge of the survey.

%G$32.0-4.9$ (3C\,396.1). Very few measurements! Possible large shell approx 60 arcmin in size. Milne \& Hill 1969 at 2.7 GHz. Very diffuse and difficult to separate background. But is visible at 2.7 and 10 GHz so possible. Is it visible at 30 GHz (maps are saturated)?

\subsection{Radio Recombination Lines}
\label{sec:results_rrls}

The ionized gas within H\textsc{ii} regions and the surrounding diffuse interstellar gas (DIG) within the warm interstellar medium (WIM) not only emit continuum free-free emission but also, when the gas is optically thin, can be observed via radio recombination lines (RRLs) \citep{Kardashev1959}. As well as an alternative probe of the WIM, RRLs provide additional velocity information, and can be used to study the physics of the interstellar medium, such as electron temperatures, abundances, ionization fractions, and non-LTE effects \citep{Brown1978}. For the study of AME at 30\,GHz, RRLs are useful because they have a direct relationship with the continuum free-free emission \citep{Rohlfs2000,Alves2015} in both discrete H\textsc{ii} regions \citep[e.g.][]{Paladini2003,Bania2010,Anderson2011} and  the DIG \citep{Luisi2020,Anderson2021} from the leakage of Lyman-$\alpha$ photons from nearby H\textsc{ii} regions \citep[e.g.,][]{Zurita2000}.

As well as measuring continuum emission from the Galactic plane, COMAP can also map emission lines between $26 < \nu < 34$\,GHz. In this first analysis we will discuss  the five hydrogen RRLs within the COMAP band: H(62)$\alpha$ to H(58)$\alpha$. The COMAP spectrometer velocity resolution is $\sim 20$\,km\,s$^{-1}$, which is just wide enough to resolve these RRLs which have a typical $\Delta V_\frac{1}{2} \sim 25$\,km\,s$^{-1}$. We will focus our attention on the bright region W43 where RRLs have been extensively studied before \citep[e.g.][]{Hoglund1965,Alves2015,Luisi2020}. 

To process the RRLs we selected chunks of spectra within $\pm 200$\,km\,s$^{-1}$ around the rest frequencies of the five available RRLs. Then for each observation we fit a first-order polynomial across the spectrum, excluding the velocity range \change{(in local standard of rest, $V_{\textrm{LSR}}$)} known to contain W43 of $0 < V_{\textrm{LSR}} < 175$\,km\,s$^{-1}$. To map the spectra we used a simple weighted averaging map-making method since, unlike with the continuum data, we can assume the data are dominated by white instrumental noise. The calibration, observations and other data processing are the same as described in \S\ref{subsec:observations}.

\begin{figure}[t] 
\begin{center}
\includegraphics[width=0.475\textwidth]{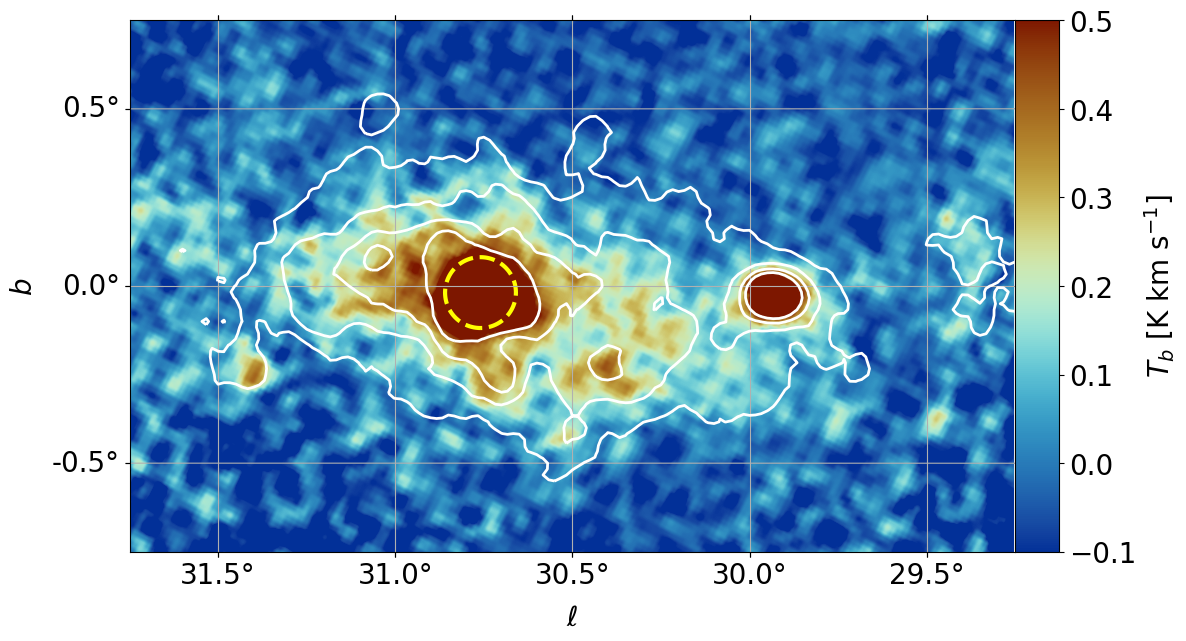}
\caption{Map of the total RRL integrated emission in units of K km\,s$^{-1}$ after stacking all five RRLs in velocity. Contours correspond to the measured continuum emission at 26.5\,GHz. The \textit{dashed-yellow} circle indicates the area used to measure the lines shown in Fig.~\ref{fig:COMAP-RRL_lines}.}
\label{fig:COMAP-RRL_map}
\end{center}
\end{figure}

In Fig.\,\ref{fig:COMAP-RRL_map} we show a map centered on W43 of the integrated RRLs after stacking all five lines; the contours are of the COMAP 26.5\,GHz continuum data. The map has units of K\,km\,s$^{-1}$ and a resolution of $4.5^{\prime}$. We can clearly see in the center of the map the W43 complex, but we also see the bright H\textsc{ii} region and molecular cloud complex G29.93$-$0.03, also known as W43-South. Between the W43 and W43-South complexes we can  see RRL emission originating in the DIG, which can be seen to be strongly correlated with the continuum emission shown by the contours. The detection of RRLs in the DIG at a similar resolution to COMAP has also recently been made at 4--8\,GHz \citep{Luisi2020,Anderson2021}.% which if combined with the COMAP data would allow for studying whether the DIG is in local thermal equilibrium. ??

\begin{figure}[t]
\begin{center}
\includegraphics[width=0.475\textwidth]{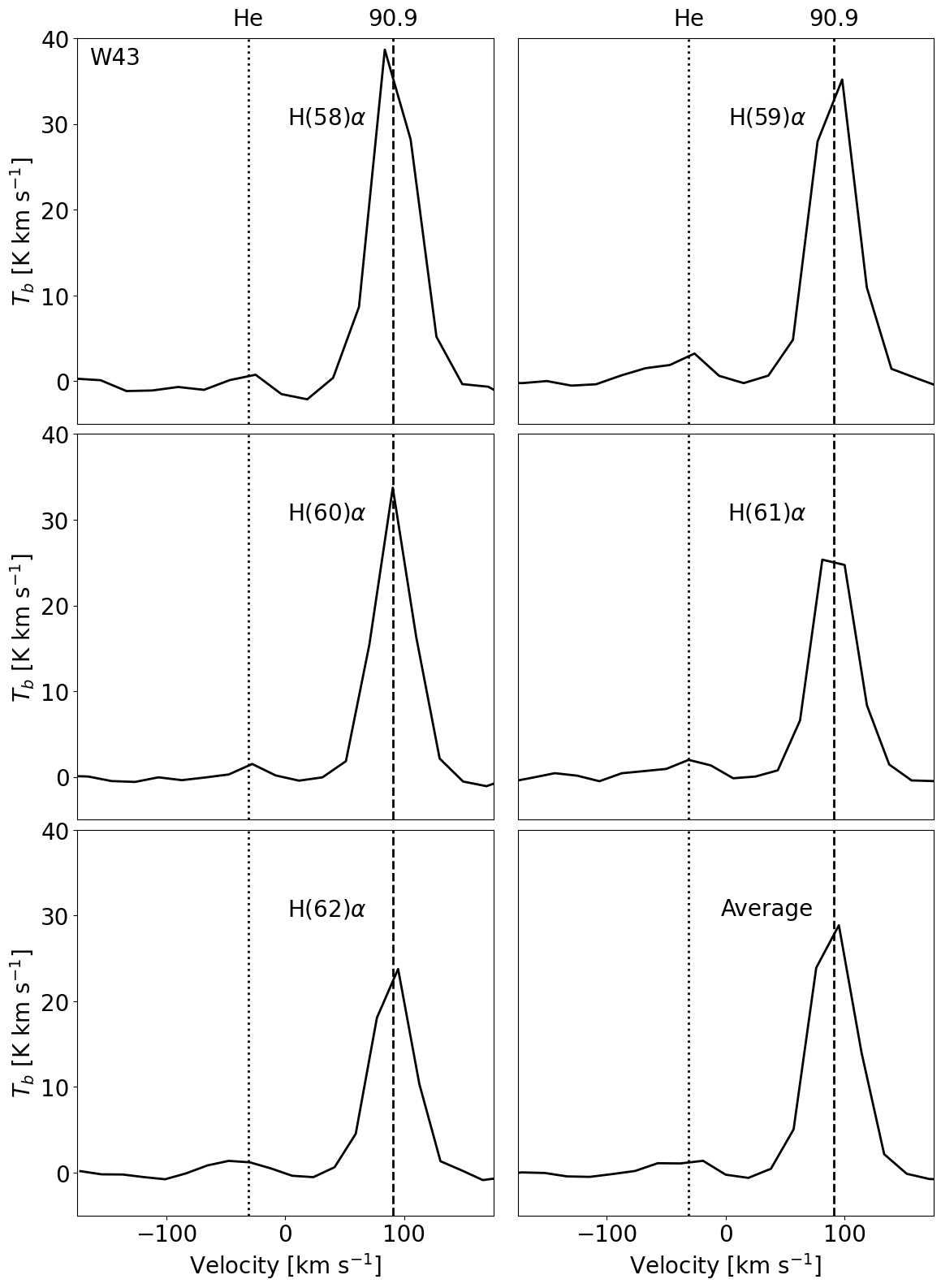}
\caption{RRL spectra for the region indicated in Fig.~\ref{fig:COMAP-RRL_map} centered on W43. We show each of the five RRLs in the COMAP 26--34\,GHz band at a resolution of $\sim20$\,km\,s$^{-1}$. The final panel shows the average of all five lines binned into a velocity bins of $20$\,km\,s$^{-1}$. The \textit{dashed-black} line marks the mean peak velocity of the five RRLs. We mark the nearby Helium RRL with the \textit{dotted-black} line.}
\label{fig:COMAP-RRL_lines}
\end{center}
\end{figure}

Figure~\ref{fig:COMAP-RRL_lines} shows the individual lines measured along the line of sight towards W43 (marked by the \textit{dashed-yellow} circle in Fig.~\ref{fig:COMAP-RRL_map}). We make clear detections of the H(58)$\alpha$ to H(62)$\alpha$ lines but we also see evidence for \change{the He transition, which is} offset from the main H transition line \change{by $-122.166$\,km\,s$^{-1}$.} The final panel of Fig.~\ref{fig:COMAP-RRL_lines} shows the average of all five lines. The fitted peak velocity of the region is found to be $V = 90.9$\,km\,s$^{-1}$, which agrees with other RRL surveys of W43 \change{\citep[$97.5\pm0.6$\,km\,s$^{-1}$,][]{Alves2012}}.

It is possible to estimate the electron temperature ($T_e$) using the known relationship between the ratio of the integrated line brightness and the underlying continuum emission

\begin{equation}\label{eqn:rrl1}
  \frac{\int T_L \mathrm{d}V_{\mathrm{km~s}^{-1}}}{T_C} = 1.0534 \times 10^{4} \frac{T_e^{-1} \nu_\mathrm{GHz}^{1}}{g_\mathrm{ff}(\nu,T_e)},
\end{equation}
where $\int T_L \mathrm{d}V$ is the integrated line emission, $T_C$ is the continuum emission brightness, $\nu$ is the frequency, and $g_\mathrm{ff}(\nu, T_e)$ is the Gaunt factor given by Equation~\ref{eqn:gaunt}.

\begin{table}
\centering
\caption{RRL properties of the W43 region  (Fig.~\ref{fig:COMAP-RRL_map}). The electron temperature ($T_e$) is calculated using Equation~\ref{eqn:rrl1}. $V$ is the fitted peak frequency of the RRL transitions, and $\frac{T_L \mathrm{d}V}{T_C}$ is the line-to-continuum brightness ratio.}\label{table:rrl}
\begin{tabular}{l|ccc}
\hline
Line& $V$ (km\,s$^{-1}$) & $\frac{T_L \mathrm{d}V}{T_C}$ (km\,s$^{-1}$) & $T_e$ (K) \\ \hline
H(62)$\alpha$ & $90.9 \pm 0.2$ & $ 8.53 \pm 0.15$  & $7810 \pm 120$ \\
H(61)$\alpha$ & $91.2 \pm 0.1$ & $ 8.67 \pm 0.10$   & $8060 \pm 80$ \\
H(60)$\alpha$ & $90.8 \pm 0.1$ & $ 9.14 \pm 0.09$  & $8080 \pm 70$ \\
H(59)$\alpha$ & $91.1 \pm 0.2$ & $10.04 \pm 0.09$ & $7840 \pm 60$ \\
H(58)$\alpha$ & $90.4 \pm 0.1$ & $ 9.70 \pm 0.07$   & $8460 \pm 50$ \\
 \hline

\end{tabular}
\end{table}

In Table~\ref{table:rrl} we report the measured line velocities, line-to-continuum ratios, and the derived electron temperatures using Equation~\ref{eqn:rrl1} for each RRL measured towards W43. We find that the derived electron temperatures are much higher than those previously reported; for example in \citet{Alves2012} the electron temperature of W43 was found to be  $T_e = 5660 \pm 190$\,K. The reason for this discrepancy is that we have not attempted to correct for contributions to the continuum emission from sources other than free-free emission, hence we are overestimating the continuum emission since there will be some contribution from AME and to a lesser extent synchrotron emission. An overestimate of the continuum brightness then leads to an overestimate of the electron temperature.

As discussed in earlier sections the two main emission components around 30\,GHz will be either free-free emission or AME. By leveraging our  knowledge of the electron temperature of W43 from \citet{Alves2012} we can  estimate what fraction of the total emission is due to AME. The fractional AME, for a given RRL, can be defined as 

\begin{equation}\label{eqn:rrl3}
  \eta_\mathrm{AME}^\mathrm{RRL} = \frac{T_\mathrm{AME}}{T_\mathrm{ff} + T_\mathrm{AME}} = 1 - \frac{\hat{a}_{n}}{a_{n}} = 1 - \frac{T_\mathrm{ff}}{T_\mathrm{ff} + T_\mathrm{AME}},
\end{equation}
where $\hat{a}_n$ is the line-to-continuum ratio we measure in Table~\ref{table:rrl} for transition H($n$)$\alpha$, $a_n$ is the expected ratio given by Equation~\ref{eqn:rrl1} and an assumed $T_e = 5660$\,K, $T_\mathrm{AME}$ is the continuum brightness of the AME, and $T_\mathrm{ff}$ is the continuum brightness of the free-free emission. We find that the mean fractional AME within W43 at 30\,GHz is $\eta_\mathrm{AME}^\mathrm{RRL}=0.35\pm0.01$, i.e., $(35 \pm 1)\%$, which is consistent with previous estimates \change{\citep[e.g. 37$\pm$5\,\%][]{Irfan2015}}.

These initial results show that the COMAP survey can readily detect RRLs from the Galaxy. The sensitivity is sufficient \change{to} detect several RRLs not only from the brightest H\textsc{ii} regions but also weaker sources including the DIG from multiple sight-lines. Although not a dedicated survey with high velocity resolution, this will be a useful dataset for the community.

\normalsize
\section{Discussion and Conclusions}
\label{sec:discussion}

We have presented the first large-scale continuum/spectral survey at 30\,GHz with an angular resolution significantly better than WMAP and {\it Planck} at the same frequencies. The COMAP instrument, being a focal plane array with a sensitive wide-band receiver, is ideal for such a survey, which will eventually cover the Galactic plane at declinations $\gtrsim -10\degree$, which covers $\ell\approx 20\degree$--$220\degree$. We will also extend the latitude range for some longitudes where necessary, such as for the W40 region at $b \approx +3.\!\!^{\circ}5$.

In this first paper, we have shown that there is sufficient S/N to measure a large number of compact sources as well as diffuse emission. Furthermore, we can reliably extract several RRLs, which \change{will provide} a unique survey in itself. The calibration is good to at least $5\%$ and potentially $\approx 1\%$ in the future. 

On large scales $\gtrsim 30^{\prime}$, there remain some issues to be resolved. First, the beam sidelobes mean that the effective calibration can change by $\approx 10\%$ \change{as a function of scale}. This can be corrected via deconvolution of the theoretical beam, although, in practice, this can be difficult due to missing pixels and accurate knowledge of the beam sidelobes. Similarly, on scales $\gtrsim 1\degree$, filtering of the time-ordered data to reduce atmospheric and instrumental $1/f$ on scales of $\approx 1\degree$ has caused a small loss of flux on these scales; our comparison of the integrated W43 flux density over a $2\degree$-diameter aperture compared to WMAP/\textit{Planck} indicates a loss of $\approx 15$--20$\%$. This filtering is also why large-scale modes away from the bright Galactic plane in Fig.\,\ref{fig:COMAP-hits+map} are not well-constrained. This will be addressed and quantified using detailed simulations and comparisons with WMAP/{\it Planck}, in future work. Our hope is that by tuning the data analysis pipeline we can reduce these effects. Alternatively, we can combine the COMAP data on scales $\lesssim 30^{\prime}$ and WMAP/{\it Planck} data on scales $\gtrsim 30^{\prime}$ to improve accuracy on all scales down to $5^{\prime}$.

In this initial work, we have shown that a wide range of science can be investigated. The 30\,GHz map is, on the whole, dominated by optically thin free-free continuum emission, with contributions from synchrotron emission along some lines of sight due to supernova remnants. We also find that there is evidence for a significant contribution of AME at 30\,GHz, both from the diffuse interstellar medium at a level of \change{$\eta_\mathrm{AME}^\mathrm{diffuse} = (22 \pm 2)$\,\%} of the total. For the individual regions we find the average percentage of AME at 30\,GHz of $\left<\eta_\mathrm{AME}^{\mathrm{H\textsc{ii}}}\right> = (43 \pm 3)$\,\%. We emphasize that this value is likely biased high (compared to all H\textsc{ii} regions) since we have chosen regions with AME and it is not a complete sample. Nevertheless, it is comparable to what has been found in some other H\textsc{ii} regions \citep{TodorovicEtAl2010,PlanckIntXV}. 

In \S\ref{sec:results_rrls} we find that RRLs under-predict the observed 30\,GHz continuum emission within the core of W43; this excess we have interpreted to be due to AME with $\eta_\mathrm{AME}^\mathrm{RRL} = (35 \pm 1)$\,\%. Interestingly in \S\ref{sec:results_hii} we find no evidence of AME from the central core of W43, apparently contradicting the AME estimated by the RRLs, but this is because the RRL estimate also includes the contribution from the large-scale diffuse background behind the core of W43---implying the AME is associated more with the diffuse ISM and not the central W43 complex in this region. 

%The apparently much higher AME fractions of the individual regions than in the diffuse emission will mostly be due to a selection bias effect i.e., sources with higher AME fractions are more likely to be flagged as detections of AME. Therefore, the individual regions are more likely to represent special cases (similar to Perseus and $\rho$\ Ophiuchi molecular clouds), while the diffuse emission fraction is likely to represent the true average AME fraction within the inner Galaxy.

Of the nine H\textsc{ii} regions discussed in \S\ref{sec:results_hii} we presented six that we find to contain AME and three that do not contain AME. As discussed above, the H\textsc{ii} regions that we find to contain AME are probably special cases, as the diffuse emission and RRLs suggest AME is everywhere at 30\,GHz. The environments of the six AME regions are diverse. PDRs have been suggested to be associated with AME \citep{Casassus2008,Arce-Tord2020} and we see evidence for PDRs in four of the regions: G027.266$+$00.165, G028.658$+$00.030, G029.038$-$00.636, and G030.443$+$0.435, but we do not see a PDR within one of our most significant AME detections (G030.508$-$00.447). Another possibility is the density of the dust within the region, because dark clouds have been found previously to be progenitor sources for AME \citep[e.g.,][]{Dickinson2010,Vidal2011,HarperEtAl2015}, but again we find that IRDCs are prominent within each region and there is no correlation with the average or maximum IRDC optical depths. Using the data from the CORNISH survey we do find that each of the AME regions contains at least one UCH\textsc{ii} region, while two of the non-AME regions contain none. However, we have shown that, in general, these UCH\textsc{ii} regions are not significant contributors to the observed emission at 30\,GHz. The presence of UCH\textsc{ii} regions is a tracer of high-mass star formation \citep[e.g.,][]{Habing1979} indicating the AME regions may have stronger local interstellar radiation fields; this is thought to have an indirect impact on AME emissivity. Indeed, numerous studies have found a correlation of the AME emissivity and interstellar radiation field strength \citep[e.g.,][]{Tibbs2012,PlanckIntXV}.

Modified blackbody fits to \textit{Planck} HFI data (217--857\,GHz) suggest that the Rayleigh--Jeans (R-J) tail of thermal emission is not significant at 30\,GHz for these regions. Nevertheless, it will be interesting to search for the densest infrared dark dust clouds \citep[e.g.,][]{Jyothish2020} in the outer galaxy (where diffuse free-free emission will be much weaker) that have yet to ionize their surroundings, where thermal dust may be detectable with the COMAP data. This would allow a test of the emissivity power-law model for thermal dust to much lower frequencies. These may also be strong AME emitters \citep{Scaife2010,Ysard2011,Tibbs2015}.

\begin{figure}[tbh]
\begin{center}
\includegraphics[width=\columnwidth]{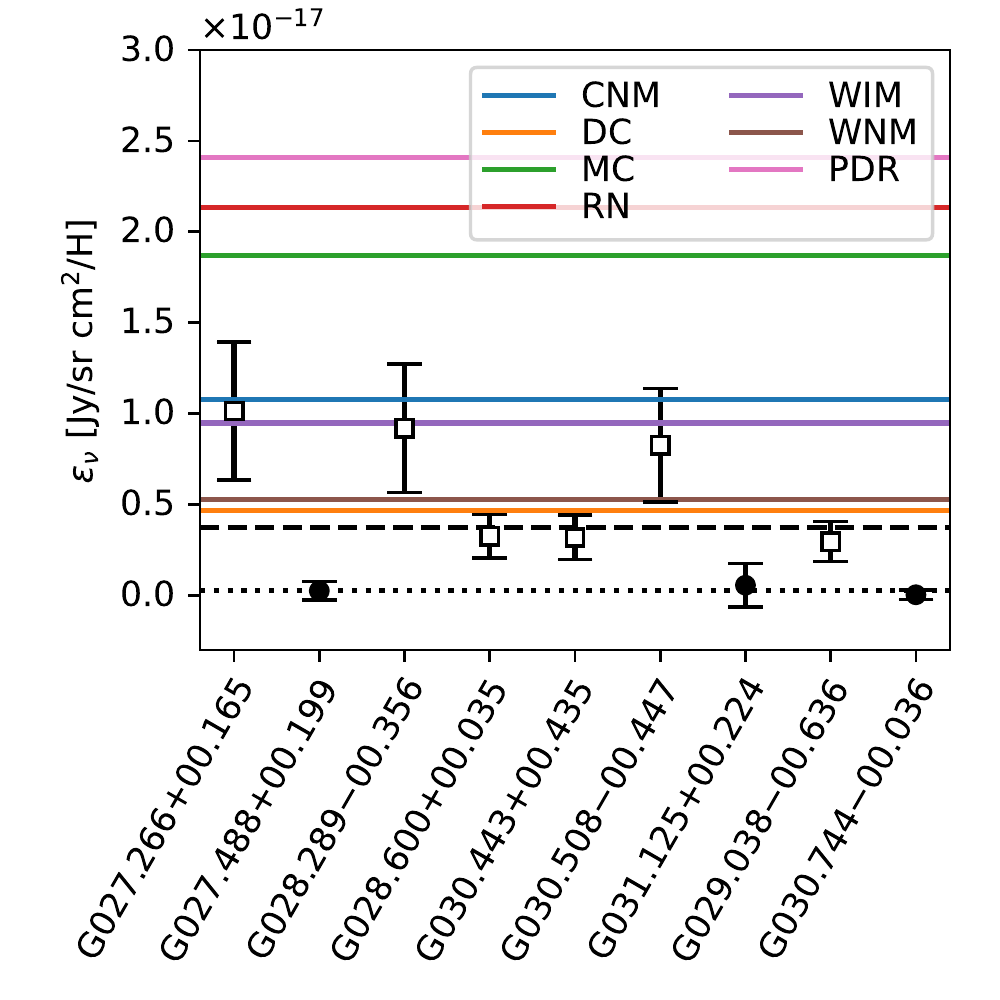}
\caption{Plot of emissivities \change{(calculated using the source sizes in Table \ref{tab:HII})} of all highlighted sources on the Galactic plane at $30$\,GHz across the different regions. The colored solid lines represent spinning dust models for a range of typical interstellar environments: cold neutral medium (CNM), dark clouds (DC), molecular clouds (MC), reflection nebulae (RN), warm ionized medium (WIM), warm neutral medium (WNM) and photodissociation region (PDR). Sources categorized as AME or non-AME regions are plotted as white squares and filled circles, respectively. The weighted average values are shown for the AME (dashed line) and non-AME (dotted line) regions.}
\label{fig:AME2}
\end{center}
\end{figure}

In Fig.~\ref{fig:AME2} we show the AME emissivities at 30\,GHz for each of the nine H\textsc{ii} regions compared with predictions from \texttt{SpDust} \citep{Ali-HaimoudEtAl2009,SilsbeeEtAl2011} for six different environments using the parameters given in \citet{DraineLazarian1998}. The weighted average emissivity of the six AME regions is $(3.7 \pm 0.1) \times 10^{-17}$\,Jy\,sr$^{-1}$\,cm$^{2}$\,H$^{-1}$ and is marked in the figure with a \textit{black-dashed} line. The uncertainties on the emissivities are driven by the uncertainty in the conversion of dust optical depth to column density as discussed in \S\ref{sec:results_hii}. We find that the six AME detections have emissivities that are of the  magnitude expected from the models. We note that the model emissivities should not be overly interpreted since the environmental parameters are only rough guides and could change by a factor of several within the same phase of the ISM. It is not clear why there are two groups of AME emissivities. We do find that in the less emissive regions there is a nearby ionizing O/B star or association \citep{Reed2003}; indeed, stronger ISRFs may destroy AME carriers \citep{Dong2011}.

We detect six out of the 33 known SNRs within the current survey region. As one would expect at this higher radio frequency, we are more sensitive to flat-spectrum ($\alpha > -0.5$) than to the majority of SNRs ($\alpha \lesssim -0.5$). These are typically filled-center or composite SNRs (filled-center and a shell). These are often classed as ``plerionic,'' containing emission from the central region as well as shells from shock waves. Typically, they host a pulsar and are referred to as ``Pulsar Wind Nebula'' (PWN), similar to the Crab nebula (Tau~A). 2 out of the 6 have $\alpha\approx-0.3$ from 2.7 to 30\,GHz while G021.5$-$0.9 remains remarkably flat ($\alpha \approx 0$) to 30\,GHz. Two out of the six have steeper spectral indices ($\alpha \approx -0.7$). At COMAP frequencies, we show evidence for steeper spectral indices indicative of spectral ageing for two sources, while for W44 we find it is consistent with extrapolations from lower frequencies, in contradiction to recent results of \cite{Loru2019}. Of the 294 SNRs known \citep{Green2009} $79\%$ are classified as shell-SNRs, with $\approx 20\%$ as filled or composite, which are more likely to be flat spectrum. Extrapolating based on these statistics and the number and level of COMAP detections thus far, we expect to detect $\sim\!50$ SNRs in the complete survey.

The spectroscopic nature of the COMAP data has also allowed us to extract five hydrogen RRLs with a spectral resolution of 20\,km\,s$^{-1}$. Although the COMAP instrument was not designed for Galactic spectroscopic science (ideally we would have much higher spectral/velocity resolution), we have shown that RRLs can be reliably extracted. The COMAP RRL survey represents the highest frequency Galactic RRL survey to-date. A comparison with the lower frequency RRL surveys from the GBT GDIGS 4--8\,GHz RRL survey \citep{Anderson2021} and HIPASS 1.4\,GHz RRL survey \citep{Alves2015} will be interesting, particularly for quantifying non-LTE effects. Even though they are very weak, we can map in RRLs not only the bright H\textsc{ii} regions (which are brightest in free-free emission) but also diffuse emission away from H\textsc{ii} regions down to a level of $\approx 0.1$\,K\,km\,s$^{-1}$. RRL data can be used to estimate electron temperatures or, if they are known, can be used to estimate the free-free continuum. Applying this to W43, we found that the continuum was $\approx 35\%$ brighter than expected, which we interpret as AME. The RRL data will be particularly useful for subtracting a model of free-free from the continuum map in order to investigate non-free-free components such as SNRs, AME, and the R-J tail of thermal dust.

Additional data will be important both for improving the SEDs and for understanding the nature of the interstellar medium. At higher longitudes ($\ell \gtrsim 60^{\circ}$) we will be able to include additional radio data from the 1.4\,GHz Canadian Galactic Plane Survey \citep{Landecker2010} and AMI 15\,GHz survey \citep{Perrott2015}. One limitation is the lack of data at frequencies $\sim 40$--100\,GHz, which would significantly help in measuring the spectral shape of the AME (peak frequency, width). This may be possible in the future with ALMA bands 1--3 covering $\approx 35$--116\,GHz, using both the total-power and interferometric modes. 

The COMAP 26--34\,GHz survey is currently on-going and is expected to be completed by 2023/24, covering the majority of the northern Galactic plane. The data, both continuum and RRL maps, will be released to the community when it is completed. In the mean time, we will be using the data to study the types of sources discussed in this paper in more detail. For example, when we have a large number of AME detections, we will be able to investigate the environments to understand what is driving the spinning dust. 

%% To help institutions obtain information on the effectiveness of their 
%% telescopes the AAS Journals has created a group of keywords for telescope 
%% facilities.
%
%% Following the acknowledgments section, use the following syntax and the
%% \facility{} or \facilities{} macros to list the keywords of facilities used 
%% in the research for the paper.  Each keyword is check against the master 
%% list during copy editing.  Individual instruments can be provided in 
%% parentheses, after the keyword, but they are not verified.

\begin{acknowledgements}

\change{This material is based upon work supported by the National Science Foundation under Grant Nos.\ 1517108, 1517288, 1517598, 1518282 and 1910999}, and by the Keck Institute for Space Studies under ``The First Billion Years: A Technical Development Program for Spectral Line Observations''.

DC is supported by a CITA/Dunlap Institute postdoctoral fellowship. The Dunlap Institute is funded through an endowment established by the David Dunlap family and the University of Toronto. CD acknowledges support from an STFC Consolidated Grant (ST/P000649/1). Work at the  University of Oslo is supported by the Research Council of Norway through grants 251328 and 274990, and from the European Research  Council (ERC) under the Horizon 2020 Research and Innovation Program (Grant agreement No. 819478, \textsc{Cosmoglobe}). JG acknowledges support from the University of Miami and is grateful to Hugh Medrano for assistance with cryostat design. SH acknowledges support from an STFC Consolidated Grant (ST/P000649/1). Isu Ravi, as appropriate. LK was supported by the European Union’s Horizon 2020 research and innovation programme under the Marie Skłodowska-Curie grant agreement No. 885990. JK is supported by a Robert A. Millikan Fellowship from Caltech. Part of this research was carried out at the Jet Propulsion Laboratory, California Institute of Technology, under a contract with the National Aeronautics and Space Administration. At JPL, we are grateful to Mary Soria for for assembly work on the amplifier modules and to Jose Velasco, Ezra Long and Jim Bowen for the use of their amplifier test facilities. HP acknowledges support from the Swiss National Science Foundation through Ambizione Grant PZ00P2\_179934. RR acknowledges support from ANID-FONDECYT grant 1181620. MV acknowledges support from NSF AST-1517598 and a seed grant from the Kavli Institute for Particle Astrophysics and Cosmology.

We would like to thank Ivayla Kalcheva for discussions on source classification in the CORNISH UCH\textsc{ii} survey.

We would also thank Eugenio Schisano for providing the Hi-GAL maps used to closely examine the sources presented in this work.

The Scientific color maps roma and romaO \citet{Crameri2018} are used in this study to prevent visual distortion of the data and exclusion of readers with color-vision deficiencies \citet{CrameriEtAl2020}.

\change{We thank the anonymous referee, whose comments and suggestions have helped to improve and clarify this manuscript.}
\end{acknowledgements}

\facilities{CO Mapping Array Project: Pathfinder}

%% Similar to \facility{}, there is the optional \software command to allow 
%% authors a place to specify which programs were used during the creation of 
%% the manuscript. Authors should list each code and include either a
%% citation or url to the code inside ()s when available.

\software{Astropy~\citep{AstropyEtAl2013,AstropyEtAl2018};
          Emcee~\citep{EMCEE};
          h5py~\citep{H5PY};
          healpy~\citep{HEALPY};
          Matplotlib~\citep{matplotlib};
          Numpy~\citep{NUMPY};
          SciPy~\citep{SCIPY};
          Source Extractor Python~\citep{BertinArnouts1996,Barbary2016}
          }

%% Appendix material should be preceded with a single \appendix command.
%% There should be a \section command for each appendix. Mark appendix
%% subsections with the same markup you use in the main body of the paper.

%% Each Appendix (indicated with \section) will be lettered A, B, C, etc.
%% The equation counter will reset when it encounters the \appendix
%% command and will number appendix equations (A1), (A2), etc. The
%% Figure and Table counter will not reset.

%\appendix
%\section{Appendix information}
%\label{app:AppendixInfo}
%Appendices

\bibliography{references.bib,zotero.bib,early_science.bib}{}
\bibliographystyle{aasjournal}

\end{document}